\documentclass[journal]{IEEEtran}

\usepackage{cite}
\usepackage{amsmath}
\usepackage{amssymb,amsfonts}
\usepackage{algorithmic}
\usepackage{graphicx}
\usepackage{textcomp}
\usepackage{xcolor}
\usepackage{lipsum} 
\usepackage{caption} 
\usepackage{booktabs} 
\usepackage{bm} 
\DeclareMathOperator*{\argmax}{arg\,max}
\usepackage{comment} 
\usepackage{subcaption}
\usepackage{float}
\usepackage{subcaption}
\usepackage{tabularx}
\usepackage{booktabs}
\usepackage{bm}
\ifCLASSINFOpdf
\else
\fi

\begin{document}

\title{Intelligent QoS aware slice resource allocation with user association parameterization for beyond 5G ORAN based architecture using DRL}

\author{Suvidha~Mhatre,
        Ferran~Adelantado,
        Kostas~Ramantas, and~Christos~Verikoukis}

\markboth{Journal of  Class Files,~Vol.~, No.~, October~2023}%
{Shell \MakeLowercase{\textit{et al.}}: Bare Demo of IEEEtran.cls for IEEE Journals}

\maketitle

\begin{abstract}
The diverse requirements of beyond 5G services increase design complexity and demand dynamic adjustments to the network parameters. This can be achieved with slicing and programmable network architectures such as the open radio access network (ORAN). It facilitates the tuning of the network components exactly to the demands of future-envisioned applications as well as intelligence at the edge of the network. Artificial intelligence (AI) has recently drawn a lot of interest for its potential to solve challenging issues in wireless communication. Due to the non-deterministic, random, and complex behavior of models and parameters involved in the process, radio resource management is one of the topics that needs to be addressed with such techniques. The study presented in this paper proposes quality of service (QoS)-aware intra-slice resource allocation that provides superior performance compared to baseline and state of the art strategies. The slice-dedicated intelligent agents learn how to handle resources at near-RT RIC level time granularities while optimizing various key performance indicators (KPIs) and meeting QoS requirements for each end user. In order to improve KPIs and system performance with various reward functions, the study discusses Markov's decision process (MDP) and deep reinforcement learning (DRL) techniques, notably deep Q network (DQN). The simulation evaluates the efficacy of the algorithm under dynamic conditions and various network characteristics. Results and analysis demonstrate the improvement in the performance of the network for enhanced mobile broadband (eMBB) and ultra-reliable low latency (URLLC) slice categories.

\end{abstract}


\IEEEpeerreviewmaketitle

\section{Introduction}
\IEEEPARstart{B}{eyond} 
5G envisioned vertical applications and use cases have diverse requirements to satisfy the quality of service (QoS) of end users. The 5G verticals can be categorized into 3rd generation partnership project (3GPP) defined use cases such as enhanced mobile broadband (eMBB), ultra-reliable low latency (URLLC), and massive machine type communication (mMTC). With various applications, such as Industry 4.0, smart cities, V2X, video streaming, online gaming, AR/VR, etc., the key performance indicators (KPIs) for the mentioned use cases can change. Such diversity in KPIs and requirements can be achieved with the use of new technology enablers. 

Furthermore, it introduces a sophisticated programmable architecture with radio access network (RAN) intelligent controllers that provide infrastructure-based abstraction of networks as well as applications performing closed loop control for RAN radio resource management (RRM). Network slicing is another important technology enabler, as it essentially allows to customize corresponding services \cite{ref1} and to adapt to changing traffic requirements \cite{slicing1, slicing2}. Due to the diversity and complexity of the criteria for future applications under the eMBB and URLLC use cases, this research focuses on these use cases and establishes a slice category for each of them separately. It is crucial to include its awareness while allocating resources to end users.  
One of the challenges in handing resources to the RAN edge domain is traffic load variation in the wireless network environment. It affects optimal resource allocation, thereby reducing resource utilization and also causing System Level Agreement (SLA) violations as well as degrading the quality of service (QoS) of the end users. Slicing can be tuned to maximize resource utilization and QoS while isolating the performance of individual slices, even under traffic uncertainty. Fine-grained resource reconfiguration has extremely high computational complexity and is a sequential problem when it comes to network slice resource configuration \cite{survey, ref2}. Another difficult aspects of radio resource management at the RAN edge domain are in terms of connectivity preservation, rate control, offloading, etc.

Moreover, due to the non-deterministic nature of wireless channel conditions, mobility, traffic, etc., as well as their inherent complexity, dynamic RRM is a challenging task. Thus, it becomes hard to construct models using conventional algorithmic approaches. On the other hand, model-free artificial intelligence (AI) techniques offer effective solutions that are gaining momentum, making them crucial candidates to optimize dynamic RRM \cite{survey}. Reinforcement learning (RL), one of the branches of AI, has been proven useful to deal with control problems. One of the most popular algorithms is deep reinfocement learning (DRL) and its variations. The optimization of resource allocation in radio access networks has been successfully accomplished with DRL algorithms, including DQN, DDPG, etc.\cite{survey}. DRL techniques have also been proposed to optimize power consumption and other aspects in different RAN architectures before ORAN, such as cloud RAN \cite{SurveyRef2}. The study presented in this paper focuses on integrating the aforementioned technological enablers and proposes deep Q-learning-based techniques to implement effective, dynamic, optimal resource management at the RAN edge domain in order to provide end users that fall into various slice categories with high-Quality of Experience (QoE) and Quality of Service (QoS).

\section{Related Work}
The ORAN Alliance is actively introducing a non-proprietary version of the RAN system that allows interoperation between network components by different vendors. In the past, intelligent solutions have also been proposed with proprietary architectures such as \cite{SurveyRef1, SurveyRef2, adaptive} but ORAN architecture introduces a sophisticated programmable approach with radio access network (RAN) intelligent controllers that provide infrastructure-based abstraction of networks. In the ORAN architecture, there are two separate components: the non-real-time RAN intelligent controller (non-RT RIC) and the near-real-time RIC (near-RT RIC), which reinforces the importance of the two control levels with different time granularities \cite{oran2}. The ORAN report and specification include a comprehensive review of a number of use cases that are expected to be integrated with architectural elements to exchange data between various network components. One of the use cases describes how to integrate intelligence at the network's edge in regards to radio resource management (RRM) \cite{oran3}. 

These components of the ORAN framework enable the integration of dynamic changes at almost real-time levels for radio resource allocation to end users. Slicing is one of the enablers that provide isolated resources. The customized network components are dedicated and isolated specific to each slice category \cite{ref1, slicing1, slicing2}. There are two different levels of resource allocation and management: intra- and inter-slice, which help manage RAN edge domain resources effectively. The authors in \cite{adaptive} define an optimization problem that aims to maximize user perceived throughput while minimizing packet delay violations by modifying the MAC scheduling algorithm's parameters. It comes under the intra-slice resource allocation category. This specific work is limited to traditional network architecture, proposing optimization for individual base stations and their underlying users. Such proprietary solutions can be used with ORAN architecture, but they lack the adaptation to utilize programmable ORAN-based infrastructure, which facilitates access to the information available at a centralized entity from other base stations or ORAN radio units (ORUs), which plays a crucial role in RRM decisions. 

Reinforcement learning (RL) is a very effective tool to achieve optimal decisions in complex environments, such as non-deterministic wireless networks, due to several aspects discussed in Section 1. Traditional table-based RL techniques are infeasible to handle large state and action spaces, as concluded in \cite{ref2}. Whereas, deep reinforcement learning overcomes these limitations, like deep Q-learning and some of its variations, such as dueling deep Q-learning (Dueling DQN) and deep deterministic policy gradient (DDPG). In the case of large and multidimensional discrete action spaces, branching architecture can be introduced in dueling Q networks, as presented in \cite{ref3}. The researchers in \cite{ref2} extend the given neural network (NN) for inherently discrete multi-dimensional and large action spaces to resolve the network slice reconfiguration problem. More computational resources are required to support such solutions. 

In state-of-the art work like \cite{UA}, researchers opt for the approach of using parallel and multiple DNN to make optimal decisions for user association with the base station. The solution proposed in \cite{UA} builds multiple DNNs and a number of additional DNNs with random input to increase exploitation. It has $K$ DNNs, $N$ random decisions, and $K+N$ functions for calculating the Q value for a network with a certain number of users. So, DRL goes through all the layers in every run. Furthermore, each DNN is trained to learn the optimal decision for the entire user association (UA) matrix.  
It requires higher computation capabilities. Instead of this heuristic-style approach, the association decision can be parameterized to learn useful information about the network such as base station or ORU-based parameters. 
.
It is possible to establish various network slice resource allocation strategies by utilizing a variety of RL approaches \cite{zhang2020, xiang2020}. Dynamic network slice distribution techniques to improve performance for 5G-based bandwidth offerings is evaluated in \cite{zhang2020}. It compares different algorithms for slice resource distribution. \cite{xiang2020} uses approximation of resources and explores subchannel allocation that impacts differently with changes in multiplexing techniques. The solutions need to be formulated in such a way that they work efficiently and provide the intended outcome, independent of variations in techniques used in the wireless communication network underneath.  

Unlike the metrics taken into account in \cite{SurveyRef1}, \cite{polese2022}, \cite{yan2019}, \cite{mei2021} and other papers, the various slices in the network have different KPI thresholds to fulfill the defined QoS requirements. This poses a challenge in terms of achieving optimal system performance that is tackled in this paper by considering the variation in metrics or KPIs for different types of services that come under various RAN slice categories. 
Hence, the network should support distinct slice KPI-based reward definitions for individual end users to evaluate different service metrics. For instance, previous research \cite{khodapanah2020framework}, \cite{schmidt2019slice} considered different types of services based on varying bit rate and delay requirements, each of which took into account a distinct set of metrics for its requirements. Motivated by this work, the research presented in this paper considers the variation of QoS thresholds and KPI metrics as well as their impact while allocating resources to various slices in the formulation. This work does not prioritize only a specific service category as discussed in the above papers, such as constant bit rate. Instead, it defines weights to evaluate metrics of different slice categories included in reward function design of intelligent agents.
Furthermore, each slice category has a separate intelligent agent that learns the importance of parameters and receives rewards based on weighted metrics and QoS thresholds for different slice categories.

The proposed work discusses a system model based on ORAN architecture to deal with intra-slice RAN RRM decisions taken in real time for eMBB and URLLC slices. The problem is formulated to achieve optimal system performance and KPIs while satisfying the QoS of individual end users. In contrast to the work talked about in \cite{adaptive}, the proposed work intends to learn the importance of UA parameters with a unique approach for slice-based beyond 5G networks with multiple ORUs. The presented research work considers resource availability at all ORUs to serve the buffered traffic at users under specific categories and allocates the resources to achieve a high QoS. As indicated in introduction, traffic uncertainty affects RAN edge domain resource allocation \cite{survey, ref2}. Hence, the formulation in this paper considers parameters reflecting network traffic load that have a significant impact on resource management. It takes into consideration the most crucial aspects of a wireless communication network in terms of end users and ORUs present in the network, as indicated in the problem formulation section. 

A resource allocation or reconfiguration problem can be written in such a way that the action space is constrained to a limited number of actions to which DQN can converge. It is also vital for intra-slice RRM in terms of time granularity of 1–10 ms for the near-RT RIC component of ORAN. 

The contributions of this paper are as follows:
\begin{itemize} 
    \item The proposed algorithms deal with the association of user-ORU and serve the traffic load at individual users. In contrast to techniques discussed in the state of the art, the algorithm and intelligent agent sit at the remote edge, near-RT RIC component of the ORAN architecture. It is completely adapted to the ORAN architecture, which facilitates dynamic configurations of resources within a smaller timescale of 1–10 ms. It benefits from centralized access to the end-user and ORU information for understanding the factors affecting the respective decisions.
    \item The DQN approach iterates through all actions, and hence the proposed formulation ensures the limited number of action spaces that impact performance, reducing the convergence time. In addition to this, as it avoids large action spaces, the desired results are achieved without additional NN architectures or DQN variants, unlike \cite{ref2, ref3}. The computation capabilities required at the edge server, processing time, and training time are reduced compared to other approaches. This is vital to achieving smaller time scales for near-real-time decisions. Furthermore, the action space defined in proposed MDP is independent of number of users in the network.
    \item Instead of learning the whole user association matrix as done in the state of the art, the proposed work distinguishes itself by parameterizing the user association. The agents learn the weights for each of the parameter crucial for UA decision. With this, the computational and memory requirements can be reduced, as can the convergence time. A problem formulation is proposed that utilizes the intelligently selected weights and concludes the actual decision. It avoids the parallel and multiple DNN approaches used in several state-of-the art works for UA.
    \item Based on how well individual users perform, each intelligent agent receives rewards. The reward function is designed to include deviation of the user performance from QoS threshold value. Further these KPIs are weighted based on the slice category. The combination of these helps to get better performance compared to baseline approaches for KPIs like throughput, delay, BER, and overall system performance.
    
\end{itemize}

\section{Network Architecture}
\begin{figure*} [htbp]
        \centering
        \includegraphics[width=13cm, height=7cm]{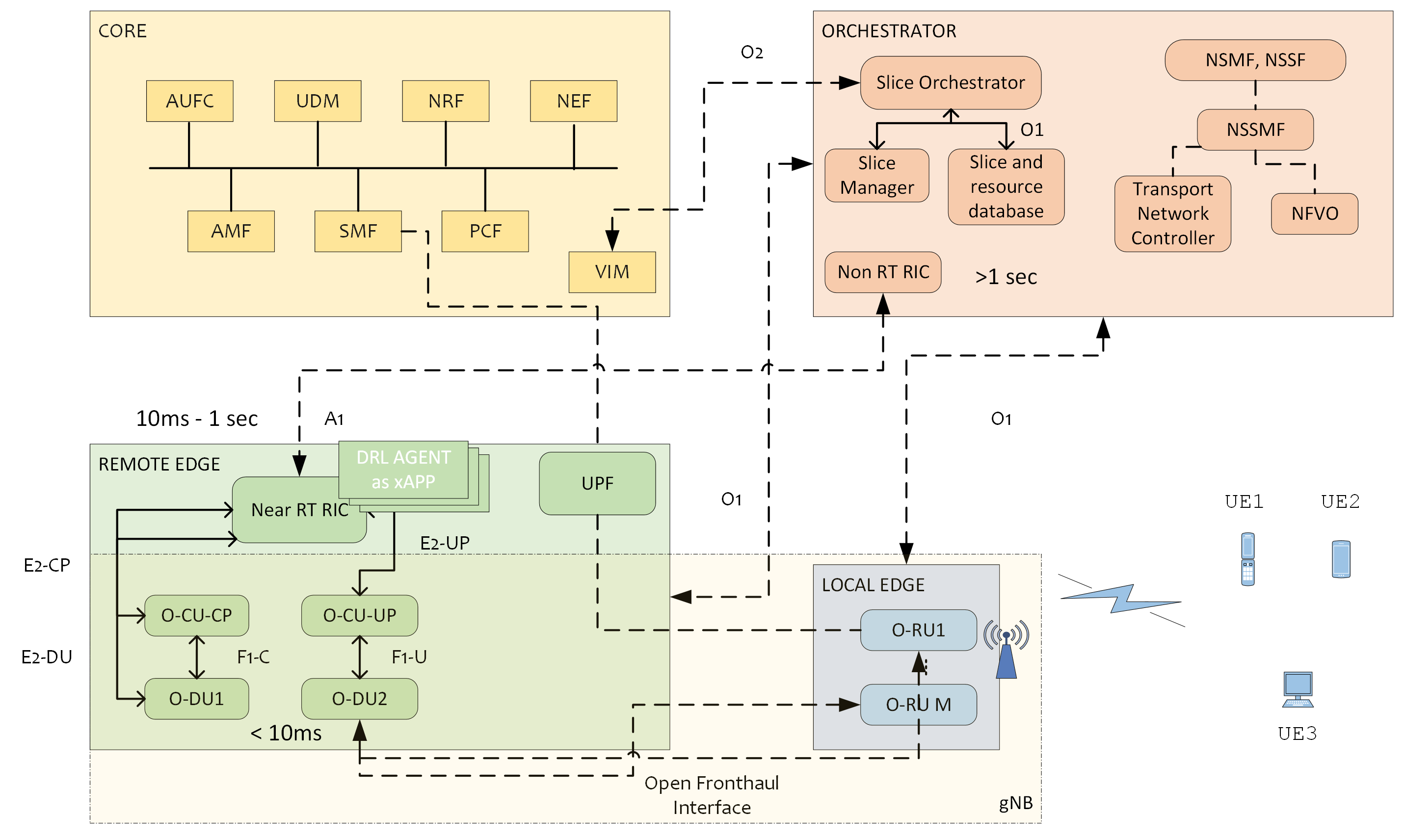}
        \captionsetup{justification=centering} 
		\caption{ORAN based Network Architecture}
		\label{fig:na}
\end{figure*}
We consider a Radio Access Network (RAN) architecture as shown in figure \ref{fig:na} based on the Open-Radio Access Network (ORAN) standard, as defined by the ORAN specifications \cite{oran1}. The RAN comprises a number of users served by various network slices, each tailored to meet specific QoS requirements. The end users are associated with the ORAN radio unit (ORU), which acts as a transceiver radio unit with antennas and a low physical layer (PHY). This unit is further connected to the ORAN distributed unit (ODU) and the ORAN centralized unit (OCU), which operate at higher protocol layers. In this RAN architecture, each user is associated with and served by an ORU under the specific Network Slice Subnet Instance (NSSI) as defined by the ORAN standard \cite{oran1}. Each NSSI corresponds to requested services and has specific KPIs. The connectivity extends to ODU, OCU, the Service Management and Orchestration (SMO) system, and the core network, providing end-to-end connectivity. The RAN Intelligent Controller (RIC) plays a crucial role in making resource allocation and management decisions intelligently at the network's edge. This increased openness allows for better decisions with more information on network parameters. 

The ORU is located at the local edge, while the ODU, OCU, and near-real-time (RT) RIC are part of the remote edge implementation at the edge servers, thus coming under multi-access edge computing. The ORU communicates with the user equipment (UE) over wireless channels, while vendor-specific open fronthaul interfaces connect the ORU to the remote edge for both uplink and downlink transmissions. For the practical implementation of the proposed algorithm, the channel state information (CSI) can be extracted from the database of the near-RT RIC at the edge server as reported by the E2 node via E2-CP. Therefore, we assume that perfect CSI is available for resource allocation and management at the edge server. Additionally, each ODU and OCU at the remote edge server connects to the SMO via O1 and A1 interfaces. The RAN communicates with core network entities via a fronthaul link, such as Ethernet, PLC, or optical fronthaul. Notably, the fronthaul link has an upper bound in terms of maximum bits transmitted per second and is subject to limited core network resources assigned to each slice or network subnet at the edge or cloud, including computational capabilities. 

The RAN NSSI resource management is executed at two different control loop levels, namely non-real-time (non-RT) and near-real-time (near-RT), in line with ORAN specifications \cite{oran1}. For each ORU, SMO provides a default RAN resource configuration via the O1 interface as part of non-RT resource management. It includes radio resources, such as bandwidth, and computational resources reserved for different types of slices. Consequently, each RAN NSSI has a pre-allocated portion of bandwidth to serve the associated users. The performance of each network slice is analyzed based on the requested QoS configurations to achieve Service Level Agreements (SLAs). Intelligent agents are used in the proposed dynamic RAN resource management policies to make sure that radio resources are given to different types of slices in the best way and meet the QoS requirements set by the slicing manager and orchestrator. Each slice is associated with a dedicated intelligent agent focused on learning decision-making parameters to serve optimal slice-specific KPIs. These agents aim to optimize individual slice performance by making radio resource management (RRM) decisions, thus operating at the near-RT RIC level.

\section{System Model}
\begin{figure*}[htbp]
    \begin{subfigure}[b]{\linewidth}
        \centering
        \includegraphics[width=\linewidth]{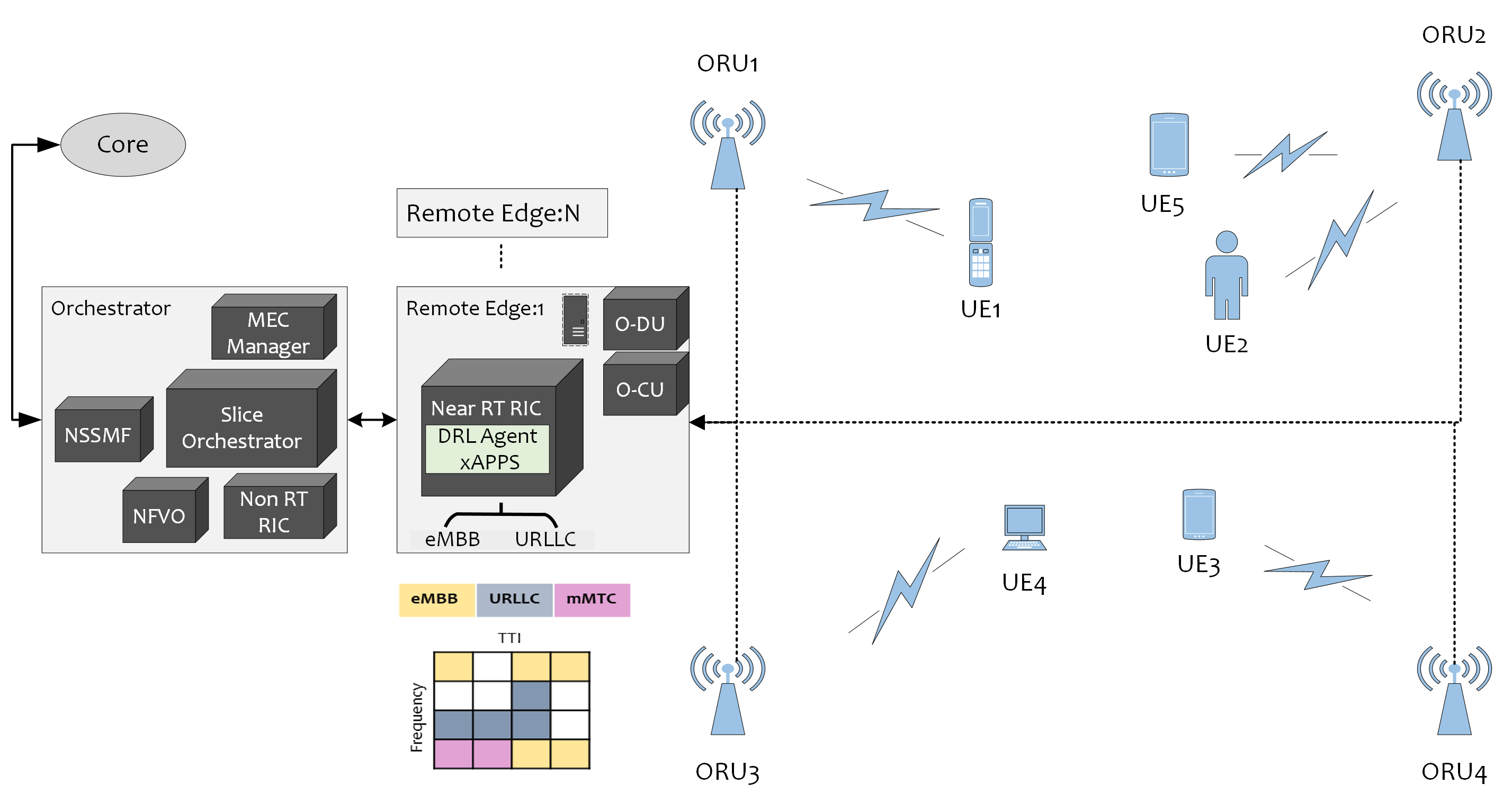} 
        \caption{System Model}
        \label{fig:subfig1}
    \end{subfigure}
    \begin{subfigure}[a]{\linewidth}
        \centering
        \includegraphics[width=\linewidth]{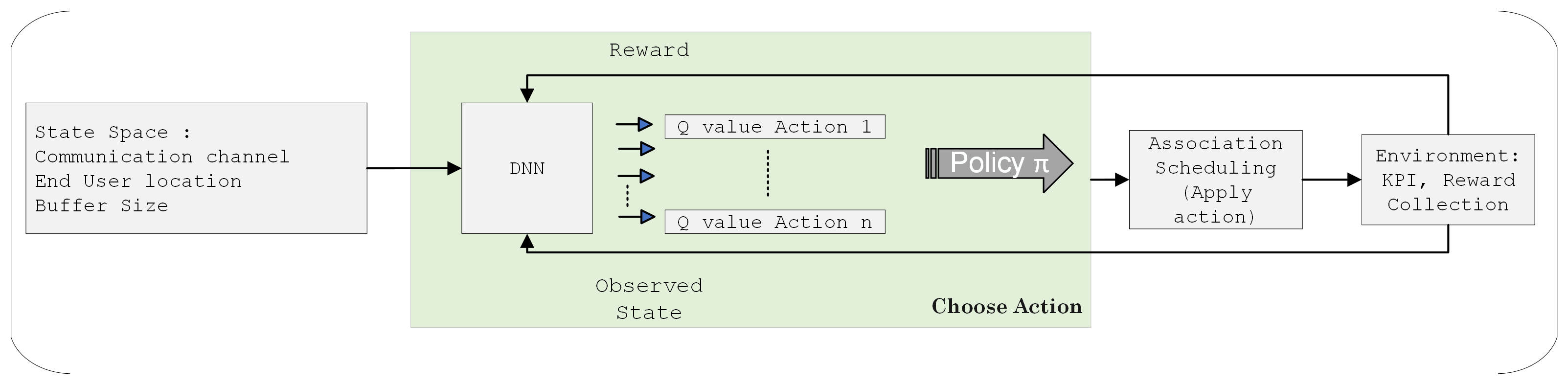} 
        \caption{DRL Agent at Near-RT RIC (Remote Edge)}
        \label{fig:subfig2}
    \end{subfigure}
    \caption{System Mdoel}
    \label{fig:carch}
\end{figure*}
As shown in the figure \ref{fig:carch}, consider the ORAN network with a general set of users $\mathcal{K} = \{ 1, \cdots, K \}$. The users can request services under any of the two slice categories eMBB and URLLC, represented by the set of users denoted as $\mathcal{K}_{E}$ and $\mathcal{K}_{U}$ respectively, where $\mathcal{K} = \mathcal{K}_{E} \cup \mathcal{K}_{U}$ and $\mathcal{K}_{E} \cap \mathcal{K}_{U} = \emptyset $. Each user has QoS requirements to be fulfilled based on the type of slice requested. These QoS requirements are expressed in terms of the minimum data rate to be achieved per user $R_k^{\text{min}}$ and the maximum allowed delay per user $d_k^{\text{max}}$ where $k \in \mathcal{K}$. We assume that all users from the same slice have the same minimum data rate and maximum delay requirements. Both the slices follow different traffic generation models due to the variation in services under these slices. The packet arrival rates for eMBB and URLLC slices follow periodic deterministic traffic model with specific packet arrival intervals. 
The size of packets and number of packets for each user vary based on the slice category and traffic distribution models, respectively. We assume it is the same for all users under the same slice category indicated for each user $k$ as $S_{k}$ and $L_{k}$, where $k \in \mathcal{K}$. The slice type as well as service requirements are mentioned for each UE in the UE database available at the remote edge. The set of ORUs denoted by $\mathcal{M} = \{ 1, \cdots, M \}$ serves these users.

To make the best use of resources within a slice, there is a separate intelligent agent for each RAN slice at the near-RT RIC. These agents learn how to better manage and assign radio resources within the ORU's assigned bandwidth. Its objective is to associate users with ORU and schedule resources in an optimal way to achieve the required QoS performance. We consider downlink (DL) frequency division duplexing (FDD) transmission. Let the binary variable $a_{k,m}$ indicate the association between ORU and users in DL and can be expressed as matrix $A$. Each user will be associated with one of the ORUs for transmission at a given time for DL. 
 \begin{equation}
     \begin{split}
         &a_{k,m} =
        \begin{cases}
            1, & \text{If } k \text{ is associated with } m \\
            0, & \text{otherwise} 
        \end{cases} 
     \end{split}  
\end{equation}
Where, $m\in \mathcal{M}, k\in \mathcal{K} $
\begin{equation}
    A=\left[\begin{matrix}a_{11}&\cdots&a_{1M}\\\vdots&\ddots&\vdots\\a_{K1}&\cdots&a_{KM}\\\end{matrix}\right]
\end{equation}

As discussed earlier, there are two types of slices: URLLC and eMBB. These slices can be instantiated in any ORU. Hence, the subset of ORUs serving $s$ slice users can be expressed as:
\begin{equation}
    \begin{split}
        &\mathcal{M}_{s} = \left\{ m \in \mathcal{M} \mid a_{k,m} = 1 \right\}, \  k \in \mathcal{K}_{s} \\ 
        & s \in \{E, U \}
    \end{split}
\end{equation}

Each slice has a separate association matrix $A_s$ with corresponding ORUs and users. The $W$ bandwidth is divided into $\mathcal{PRB} = \{ 1, \cdots, PRB \}$ sub-channels. It is also assumed that a user can only be connected to one ORU at a time, and a user may be assigned one or multiple physical resource blocks (PRBs) based on the traffic at each user. The bandwidth allocated to each ORU is divided among the slices instantiated in the ORU. Hence, ORU has a specific number of resources $PRB_m$ available, and the corresponding slice has a total $PRB_{m, s}$ number of PRBs available where $s \in \{E, U \}$. We assume additive white gaussian noise for all users of independent circular symmetric natured complex random variables with a zero mean and $\sigma^2$ variance. All users in the network follow random mobility model with speed $V$. As per 5G new radio (NR), we only consider numerology $\mu=0$. 

Here, all the parameters indicated are with respect to a single TTI t; hence, for simplicity, the notation t is omitted from all variables unless the time duration is other than 1 TTI or a reference to a future or past TTI value is required. In DL, we use $h_{k,m}$ to denote the channel coefficient from ORU $m$ to user $k$. The frequency-selective flat fading channel passes through Rayleigh fading in the wireless medium. Let us assume that $P_m$ is the total transmit power of ORU. 
Then the SINR of the $k$-th user for the wireless link with the $m$-th ORU for DL can be expressed as,
\begin{equation}
    \gamma_{k,m}=\frac{P_m\left|h_{k,m}\right|^2}{I_{k,m}+\sigma^2} 
\end{equation}
Where the interference from the $i$ ORUs (other than the associated $m$-th ORU) for the $k$-th user can be formulated as,
\begin{equation}
    I_{k,m}=\sum_{i\in{\mathcal{M},i\neq m}}{P_i\left|h_{k,i}\right|^2}   
\end{equation}

\section{Problem Formulation}
The proposed formulation aims to allocate and manage radio resources within each slice in a dynamic and optimal way such that all users satisfy the QoS requirements of requested services. The majority of existing proprietary and state-of-the-art based solutions address the importance of including intelligence at the inter-slice level. But it is equally crucial to include such intelligence in the intra-slice resource allocation to adapt to the diverse KPIs of future envisioned applications. The proposed work tries to fill these gaps while achieving optimal system performance. The proposed algorithms deal with near-RT RIC-level time granularity and are available as xAPPs at the remote edge of the network architecture. The algorithms are implemented in ORAN architectures (in the direction of cell-less), enabling dynamic radio resource allocation decisions towards end users within the available resources of each slice. It focuses on the optimal and intelligent way to learn the importance of various decision-making parameters. The formulated solutions tend to rely on parameters such as signal-to-interference noise ratio (SINR), estimation of PRBs assigned to each user, and ORU capabilities to serve the traffic load in the current network at any given time. 

The SINR values $\gamma_{k,m}$ are calculated as per equation (4) and they include effects such as path loss and shadowing based on the distance as well as the wireless channel quality between the user and ORU. Furthermore, it considers other interfering components in the network that affect the signal strength of the actual signal. The estimation of assigned PRBs is included in one of the proposed algorithms, as prior knowledge of this parameter increases the KPI performance observed through several tests. The ability of ORU to serve traffic in a network is evaluated with a performance metric for ORU, $\tau_{m}$ defined as the ratio of buffer size in a given TTI $t$ at ORU $m$ to actual transmitted bits in the same TTI for the respective ORU. This metric is a type of queueing delay estimator that indicates how fast an ORU can serve the user. 
The metric $\tau_{m}$ is calculated as below:
\begin{equation}
    \tau_{m} = \frac{C_{m}}{J_{m}} 
\end{equation}
\begin{equation}
    \begin{split}
        C_m &= \sum_{k \in \mathcal{K}_m} B_{k} \\
        J_m &= \sum_{k \in \mathcal{K}_m} TB_{k}
    \end{split}
\end{equation}
where $\mathcal{K}_m$ is a set of users associated with $m$-th ORU, $B_{k}$ is number of bits in buffer for each associated user and $TB_{k}$ is number of bits transmitted for each associated user under ORU $m$ at TTI $t$. Further, from the experienced value of $\tau_{m,t}$ in each TTI for each m-th ORU, we calculated global value $\tau^g_{m,t}$. Initially, the global value is set to zero and updated with each TTI. Each experienced $\tau_{m,t}$ within each TTI is utilized to update the global ${\tau^g}_{m,t}$ as shown below.
\begin{equation}
    \tau^g_{m,t} = \frac{ {\tau_{m,t}} + \tau^g_{m,t-1}}{2} \\
\end{equation} 
These parameters represent crucial aspects to be considered in the decision-making process of associating users with ORUs. The proposed formulation computes the indicated parameters to evaluate their importance and learn optimal decisions for the given state of the network environment.
\subsection{Intelligent QoS aware Resource Allocation (IQRA)}
The IQRA focuses on selecting the optimal association decision based on a combination of the above-discussed parameters calculated in current TTI. It sits at near-RT RIC as an xAPP. Initially, a database is generated that stores all possible combinations of associations between the available users and ORUs of a specific slice. With this database as a reference, the algorithm can calculate and estimate the required parameters that contribute to association and scheduling decisions within the available slice resources. As discussed in the earlier section, we indicate association with matrix $A$ defined in equation (2). For a simplified approach in IQRA, the association is expressed in a vector form deducted from the same matrix $A$. Each vector element is the value of the associated ORU $m$, where the index of the vector element indicates the UE $k$. It is defined as given below. This definition reduces the number of possible combinations for associating UEs with ORUs.
 
\begin{equation}
     \vec{A} = \{m_1, m_2......., m_{k}\}
\end{equation} 
For given $\mathcal{K}$ and $\mathcal{M}$, there are $I$ such combinations available. Each $i$-th combination represents a unique $A$ matrix with values of $a_{k,m,i}$ for $k,m \in \mathcal{K,M}$ as per equation (2). For each of these combinations, SINR $\gamma_{k,m,i}$ is calculated as per equation (4). Based on the range of calculated $\gamma_{k,m,i}$ value in dBm, the modulation and coding scheme (MCS) such as QPSK, 16 QAM, 64 QAM, etc. is selected for each user from the look-up table (LUT). The MCS scheme determines the modulation order ${O}_{k,i}$ for each user in every $i$-th combination and the number of bits to be transmitted per resource element $B_{{\rm RE}_{k,i}}$.

As per 5G NR configurations, the number of symbols per slot, number of slots per subframe and total number of subcarriers can be determined. Hence, we can calculate bits per PRB for each user as $B_{{\rm PRB}_{k,i}}$.

From all these parameters, the required number of PRBs at user $k$ can be estimated as follows using available number of bits in buffer at user $k$, $B_k$ and the number of bits that can be transmitted at user $k$ in each PRB $B_{{\rm PRB}_{k,i}}$,
\begin{equation}
    PRB^{\rm required}_{k,i}=\lceil \frac{B_k}{B_{{\rm PRB}_{k,i}}} \rceil
\end{equation}
Further, scheduling is performed based on the required number of PRBs for each user and the available bandwidth for each ORU. The proposed user association selection scheme is independent of the selected scheduler. In Section VIII, Proportional Fair (PF) scheduler is simulated. 
 
After scheduling, the estimate of assigned resources for each user is available: $PRB^{\rm assigned}_{k,i}$. Correspondingly, the buffer size in bits, $C_{m,i}$ along with the number of bits transmitted, $J_{m,i}$ is evaluated as given in equation (7), 
providing us the metric $\tau_{m,i}$ at each ORU as given in equation (6). The global value for each i is updated as follows: 

\begin{equation}
     \tau^g_{m,i} = \frac{ \frac{C_{m,i}}{J_{m,i}} + \tau^g_{m, t-1}}{2}
\end{equation}
The estimation of $PRB^{\rm assigned}_{k,i}$ and $\tau^g_{m,i}$ for each $i$-th association contributes to the association decision AD as given below:
\begin{equation}
    {AD}_{s,i}=w_{s1} \sum_{k,m} a_{k,m,i} \cdot PRB^{\rm assigned}_{k,i} - w_{s2} \sum_{m}   a_{k,m,i} \cdot \tau^g_{m,i} 
\end{equation}
The values for $w_{s1}, w_{s2}$ are learned and updated as described in Section 6 by intelligent agent for each slice where $s\in \{E,U\}$. The optimal association decision is selected from all the available $i$ combinations in the set with the following equation:
\begin{equation}
    AD_{s,opt} = \argmax_{i} \left({AD}_{s,i}\right)
\end{equation}
subjected to:
\begin{equation}
    \begin{split}
        &C1: d_k\le{\ d}_k^{max},\ k\in \mathcal{K} \\
        &C2: R_k\geq R_k^{min} ,\ k\in \mathcal{K}\\
        &C3: \sum_{m\in \mathcal{M}}\sum_{k\in \mathcal{K}} R_{k,m}\le N_{fronthaul} \\
        &C4: a_{k,m}\in\left\{0,1\right\},\ \forall_{k,m}\in \mathcal{K,M} \\
        &C5: \sum_{k}{PRB^{\rm assigned}_k\le{PRB}_{s}} ,\ k\in \mathcal{K}_s, \ s\in\{E, U\} 
    \end{split}
\end{equation}
Once $AD_{s,opt}$ is selected, we calculate $\tau^g_{m, t}$ using equation (8) where $\tau_{m,t}=\tau_{m,opt}$.

\subsection{Low Complexity Intelligent QoS aware Resource Allocation (LIQRA)}

This algorithm is proposed for the same objective in intra-slice RAN RRM with reduced computational complexity, which is crucial for some applications. The algorithm is constructed similarly, but evaluates association decisions in a different way. It takes association decisions individually for each user $k$ and estimates the matrix $A$. The algorithm aims to make an intelligent association decision based on a combination of signal-to-noise ratio (SNR) and the global ORU metric $\tau^g_{m}$. Here, we consider the same association matrix $A$ defined in equation (2). The signal-to-noise ratio $\zeta_{k,m}$ for user $k$ with each ORU $m$ for the given frequency flat fading channel $h_{k,m}$ is calculated as follows: 
\begin{equation}
    \zeta_{k,m} = \frac{P_m \times \left|h_{k,m}\right|^2}{\sigma^2} 
\end{equation}
The global ORU metric $\tau^g_{m}$ is calculated as discussed earlier according to equation (8). 
For each individual user $k$, the combination of signal-to-noise ratio $\zeta_{k,m}$ and metric $\tau^g_{k,m}$ with every available ORU $m$ is calculated. The algorithm evaluates the optimal association decision with the help of the following equation, where for each $k$, we check, 
\begin{equation}
    m* = \argmax_m  ( w_{s1} \cdot \zeta_{k,m} - w_{s2} \cdot \tau^g_{k,m} )
\end{equation}
The values for $w_{s1}, w_{s2}$ are selected by an intelligent agent for each slice. Once $m*$ for each corresponding $k$ is estimated, we can write the association matrix $A$ as,
\begin{equation}
    \begin{split}
         A &= \left[a_{k,m} \right]  \\
        \text{Where,} \ \ a_{k,m} &=
        \begin{cases}
            1, & \text{If } m = m*\\
            0, & \text{otherwise} 
        \end{cases} 
    \end{split}
\end{equation}   
Once the association decision is taken and matrix $A$ is estimated, SINR is calculated based on equation (4) and hence, MCS is selected. Further, the required PRBs and other parameters for each user $k$ are calculated as indicated in the earlier algorithm to schedule radio resources for end users.

\subsection{Key Performance Indicators}
Once the resources are allocated, the transmission takes place, and experienced key performance indicators such as achieved throughput, delay, and bit error rate (BER) are collected. These KPIs are stored, and the mean performance of the last TTIs for each user is forwarded to DRL. Additional KPIs, such as successful packet transmissions and packet drop rate, for each UE are also evaluated over the defined time interval. The throughput of user $k$ is calculated by multiplying the assigned number of PRBs with bits per PRB selected for transmission in that TTI. The total delay experienced per user is a combination of delays experienced by each packet in queue, transmission, and processing, as given below, where sub-index $l$ is packet index for the user.
\begin{equation}
    {d}_{k,l}=\ d_{k,l}^{tx}+d_{k,l}^{que}+d_{k,l}^{processing}   
\end{equation}
We consider the processing delay for each packet to be $2\times t_{symb}$ and the symbol duration $t_{symb}$ is calculated based on the selected configuration. Whereas, $d_{k,l}^{tx}$ for every packet per user can be calculated as the time required to transmit the number of bits in the packet, which can be a multiple of TTI based on how many TTIs the packet requires to transmit successfully. Whereas, $d_{k,l}^{que}$ is defined as the wait time in the queue or buffer at ORU before transmission. It is calculated as the difference between the transmission time stamp $t_{k,l}^{current}$ and packet arrival time stamp $_{k,l}^{arrival}$ for each packet per user. Now the delay per user k ${\bar{d}}_k$ is calculated as the averaged delay experienced by all packets $l$ given as,
\begin{equation}
    {\bar{d}}_k=\ \frac{\sum_{l\ \in L} d_{k,l}}{\left|L\right|} 
\end{equation}
The BER 
for each user $k$ associated with $m$-th ORU in every $t$ TTI for frequency selective flat fading channel is calculated as given in \cite{BERform}. The packet loss rate for every $T$ TTIs is calculated as the ratio of packets discarded in given time interval to the total number of packets transmitted for each user. The packet is discarded if it exceeds the threshold for the maximum delay limit $d_k^{max}$ mentioned in the QoS requirement for each user $k$. The successful transmission rate for each user for $T$ TTIs is calculated as the ratio of successfully transmitted packets in a given time interval to the total transmitted packet for each user in the same time interval.

\section{Proposed DRL based intelligent agents}

In this study, distinct intelligent agents for the slice types, eMBB and URLLC, are proposed. The UA decision is parameterized as discussed in above section. The DRL technique namely DQN is used by the intelligent agent to discover the significance of either the assigned PRB estimate or the SNR metric $\gamma$ and the ORU metric $\tau$ for suggested strategies. Based on the state of the network environment at any given time, the deep neural network (DNN) determines the weights for each of these parameters and evaluates the optimum course of action. The DNN trains and learns the weight values for various network environment states and scenarios. IQRA and LIQRA are then given with these values for $w_{s1}$ and $w_{s2}$ separately for each slice. The formulated solutions in the earlier section take into account the weights and evaluate a decision for ORU selection to serve the corresponding end user. This results in better performance for the KPIs compared to baseline and state-of-the-art solutions. The proposed Markov's decision process (MDP) establishes the significance of the weights for UA parameters using DQN based intelligent agents. Each intelligent agent learns different values based on the impact of individual experienced user KPIs under each slice category.

\subsection{Markov Decision Process}
The Markov Decision Process (MDP) for the intelligent agents based on the above formulation is defined with a tuple of $\left\{S_s,A_s,R_s\ ,\Gamma_s\right\}$ corresponding to state, action, reward, and discount factor, where subscript s indicates slice type for intelligent agents $s\in\{E,U\}$. The state space $S_s$ of the environment includes the current log normalized channel matrix $H_{M\times K}$, number of packets to be transmitted in the buffer per user ${\bar{P}}_k$, and the distance between each user and ORU in the current TTI t $D_{M\times K}$. 
The state space definition remains the same for the different slices, as it represents the current state of the environment based on which the actions will be chosen by the intelligent agent. Different intelligent agents learn the importance of the UA parameters, which are specific and valid to the varying traffic flow for the services and user-specific to each slice.
\begin{equation}
    \begin{split}
        &S_s =\left\{H_{M\times K},{\bar{P}}_k,D_{M\times K}\right\}  \ \text{for} \ k, m\in \mathcal{K}_s, \mathcal{M}_s\\
        &s \in \{E,U\}
    \end{split}
\end{equation}
The action space is combination of weights $w_{s1}$ and $w_{s2}$, where $w_{s1}$ and $w_{s2}$ are defined as discrete spaces as follows:
\begin{equation}
    w_{s1}, w_{s2} \in\left\{\frac{1}{N},\frac{2}{N},\ldots..,\ \frac{N}{N}\ \right\}   
\end{equation}
Hence the action space is given as,
\begin{equation}
    \begin{split}
        A_s &= (w_{s1}, w_{s2}) \ \ \text{i.e.}\\
        A_s &= \left\{\frac{1}{N},\frac{2}{N},\ldots..,\ \frac{N}{N}\ \right\} \times \left\{\frac{1}{N},\frac{2}{N},\ldots..,\ \frac{N}{N}\ \right\}
    \end{split}
\end{equation}
As defined above, the number of actions is limited, and the agent learns the importance of each parameter to make an optimal decision for each slice separately. Several deep learning techniques and MDP designs come across issues while achieving convergence with a huge action space. Whereas, both algorithms in the proposed work use

The intelligent agents will learn the optimal way to associate user-ORU and eventually schedule the resources based on observation at a given point in time in an environment of wireless communication. The reward function is designed in such a way that the agent receives the reward for the user with a positive value equal to a fraction of exceeding QoS thresholds. Whereas, if any user fails to reach the QoS threshold, it is awarded a negative value equal to the amount of KPI that failed to reach the threshold. This reward is further normalized by the threshold values themselves. The total reward is the accumulated reward of each individual user divided by the total number of users. Both IQRA and LIQRA can use this definition of reward function. The reward function can be expressed as follows:
\begin{equation}
    \begin{split}
        &{R}^i_s\left(S_s,A_s\right) = \frac{\left(\alpha_s \cdot\sum_{k}{\frac{R_{k,i}-R_k^{min}}{R_k^{min}}}+ \beta_s \cdot\sum_{k}{\frac{d_k^{max}-d_{k,i}}{d_k^{max}}} \right)}{K} \\
        &k \in \mathcal{K}_s, \ s \in\{E,U\}  
    \end{split} 
\end{equation}
Where $i$ indicates the episode or run of the DRL algorithm and $\alpha_s=1-\beta_s$. The values of $\alpha_s$ and $\beta_s$ are used to prioritize a specific KPI value according to slice type. For each agent, users are rewarded based on a prioritized KPI reward function. For example, for eMBB slice services such as social media, messaging, large file downloading, web browsing, etc., minimum data rate requirements are comparatively more stringent or prioritized than the experienced delays. Whereas for URLLC slice services, the latency or delay is prioritized as a KPI. We can define slices dedicated to specific service and set the priorities of KPIs in reward function according to the requirements of that service. This ensures that algorithms achieve optimal system performance by keeping the service requirements in check and the quality of experience of each user maintained. Once the agent learns optimal weights for different states, the slices obtain desired and better individual user and system performance. 
 
The conducted research has analyzed various designs of reward functions to get better convergence and KPIs. We use deep Q learning methods from DRL algorithms. Here, the Q value, target Q value, and loss function are calculated as given below.
\begin{equation}
    L_s=\left[{TD}_s-\left(Q^i_s\left(s_s,a_s\right)\right)^2\right]
\end{equation}
\begin{equation}
    \begin{split}
         Q^i_s\left(s_s,a_s\right) &=\ Q^i_s\left(s_s,a_s\right)+\Omega_s \\
         &\left[{R}^{i+1}_s+\Gamma_s\max_{\substack{a_s\in A_s}} {Q^{i+1}_s\left(s_s,a_s\right)}-{Q^i}^2_s\left(s_s,a_s\right)\right]\\
         \text{i.e.} \ \ Q^i_s\left(s_s,a_s\right) &=\ Q^i_s\left(s_s,a_s\right)+\Omega_s\times L_s
    \end{split}
\end{equation}
Here, $s \in \{E,U\}$ and the values for learning rate $\Omega_s$, discount factor $\Gamma_s$ as well as other hyperparameters are indicated in the simulation setup. The temporal difference, or target Q values, is calculated as follows:
\begin{equation}
    {TD}_s={r_s}^{i+1}+\Gamma_s\ \max_{\substack{a_s\in A_s}} {Q^{i+1}_s\left(s_s,a_s\right)}
\end{equation}

\section{Simulation Setup}
The designed simulator is 3GPP-standard-compliant for wireless communication models. The network elements are defined and implemented in line with the ORAN architecture. The simulation set-up is based on ITU-R M recommendations for testing and verification. It is implemented in a Python environment. It uses OpenAI-gym, Keras, and Tensorflow modules to implement DQN-based agents. The intelligent DQN agents for both slices are tested and tuned by running multiple combinations of hyperparameters. The final selected values are mentioned in Table \ref{tab:drl}. 
\begin{table}[htbp]
\centering
\caption{DQN Agent Hyperparameter}
\label{tab:drl}
\begin{tabular}{l l }
\toprule
Parameter  & Value \\
\midrule
Network Architecture & $256 \times 256$\\
Learning Rate  & $1e^{-3}$ \\
Batch Size &  64 \\
Epsilon minimum  &  0.01 \\
Epsilon decrement  & 0.99 \\
Target network update frequency  & 100 \\
Discount factor & 0.995\\
Activation & ReLU \\ 
Optimizer & Adam \\
\bottomrule
\end{tabular}
\end{table}
For each network slice, the number of users requested to be served and their QoS requirements are available in the database at near-real-time RIC. We have tested the proposed algorithms for a number of settings of QoS values based on different packet arrival rate (PAR). The simulation set up include packet arrival based on distribution models as well as constant bit rate (CBR). This variation in traffic arrival is considered to analyze the impact and verify the performance of algorithms with high, medium, and low traffic volumes and different distribution techniques. The number of packets available to transmit for each user is determined per TTI based on the mentioned models with different times of arrival in reference to the current time stamp. Similarly, for all UEs, the location update is calculated for each TTI based on the mobility model with a fixed 30 kmph speed for all users. 

As we have discussed in the system model, we assume 5G NR numerology 0 for all users, which corresponds to subcarrier spacing of 15 KHz and 12 subcarriers per PRB, hence the bandwidth of a single PRB ${W}_{PRB}$ is 180 KHz. Here, $W_{m,s}$ bandwidth is allocated to each ORU under each slice category where it is serving both eMBB and URLLC slice users, and then the number of PRBs available at each ORU is given with $W_{m,s}/{W}_{PRB}$ as shown in table \ref{tab:param_table}. The symbol duration $t_{symb}$, number of symbols per slot, and other parameters are calculated based on a selected combination of numerology and subcarrier spacing. The pathloss and shadowing models are defined as a combination of dense urban and hotspot models given in table A1-5 of ITU-R M.2412-0 \cite{ref4} and other communication model parameters are listed in Table \ref{tab:param_table}. The ITU-R M specification \cite{ref4} specifies the guidelines for the evaluation of radio interface technologies, specifically for simulation and testing purposes. The ORUs are placed around 50 meters away from each other, with their heights 3-3.5 meters and maximum power levels are 200 mW (linear power of ORU for 24.25–27.5 GHz) according to A1-16 and A1-23 of \cite{ref4} compliant with ORAN standardization and the ITU framework to support network slicing \cite{power}.

\begin{table}[htbp]
\centering
\caption{Simulation Parameters}
\label{tab:param_table}
\begin{tabular}{p{3.3cm} p{4.5cm}} 
\toprule
Parameter & Value \\
\midrule
Transmit power of ORU $P_m$ & 200mW \\
Noise variance $\sigma^2$ & -146.424 dBm \\
Noise Figure & 5 \\ 
Boltzmann constant k & 1.38e-23 J/K\\
Temperature in Kelvin (K) & 290 \\
Path Loss Exponent & 3.5\\
PRB bandwidth & 180 KHz \\
5G Numerology  & $\mu = 0$ \\
Subcarrier Spacing &  15KHz \\
No. of PRBs & $PRB_{E}$=32, $PRB_{U}$=15 \\
No. of ORUs   & 4   \\
No. of users: eMBB  & 5 \\
No. of users: URLLC & 5 \\
$\bm{R}_k^{\text{min}}$ (Mbps)& E = 16, U = 3.8 \\
$\bm{d}_k^{\text{max}}$ (ms)& E = 10, U = 2  \\    
Packet Size (Bytes) & E= 1024, U = 480 \\
Traffic Model & E = periodic deterministic with packet arrival interval 0.5 ms\\
&  U = periodic deterministic with packet arrival interval 1 ms\\

\bottomrule
\end{tabular}
\end{table}

\section{Results and Analysis}

\begin{figure}[htbp]
    \centering
    
    \begin{subfigure}{0.46\textwidth}
        \centering
        \includegraphics[width=\textwidth]{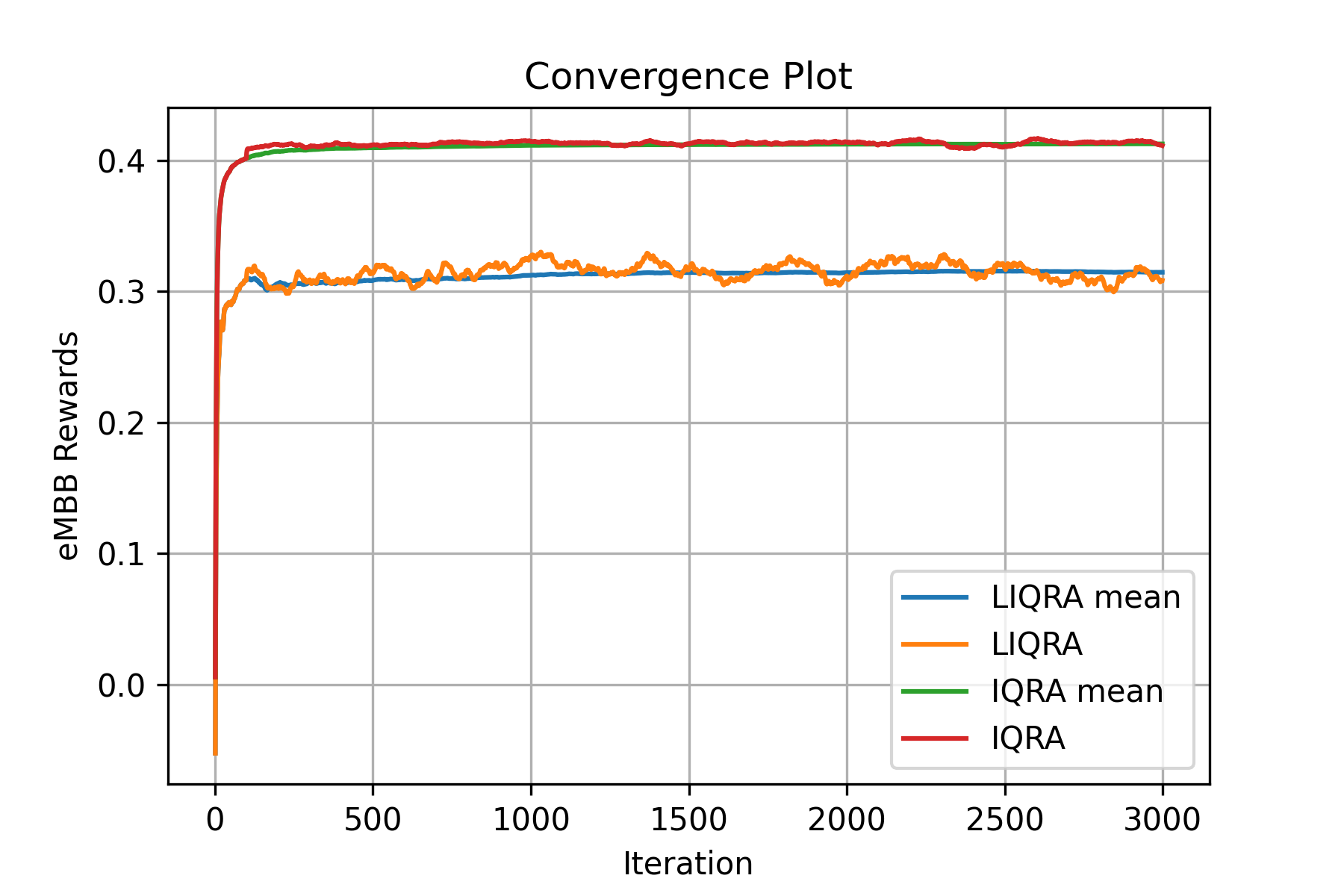}
        \caption{eMBB reward performance for DRL Intelligent agent} %
        \label{subfig:subfigure1}
    \end{subfigure}
    \hfill
    \begin{subfigure}{0.46\textwidth}
        \centering
        \includegraphics[width=\textwidth]{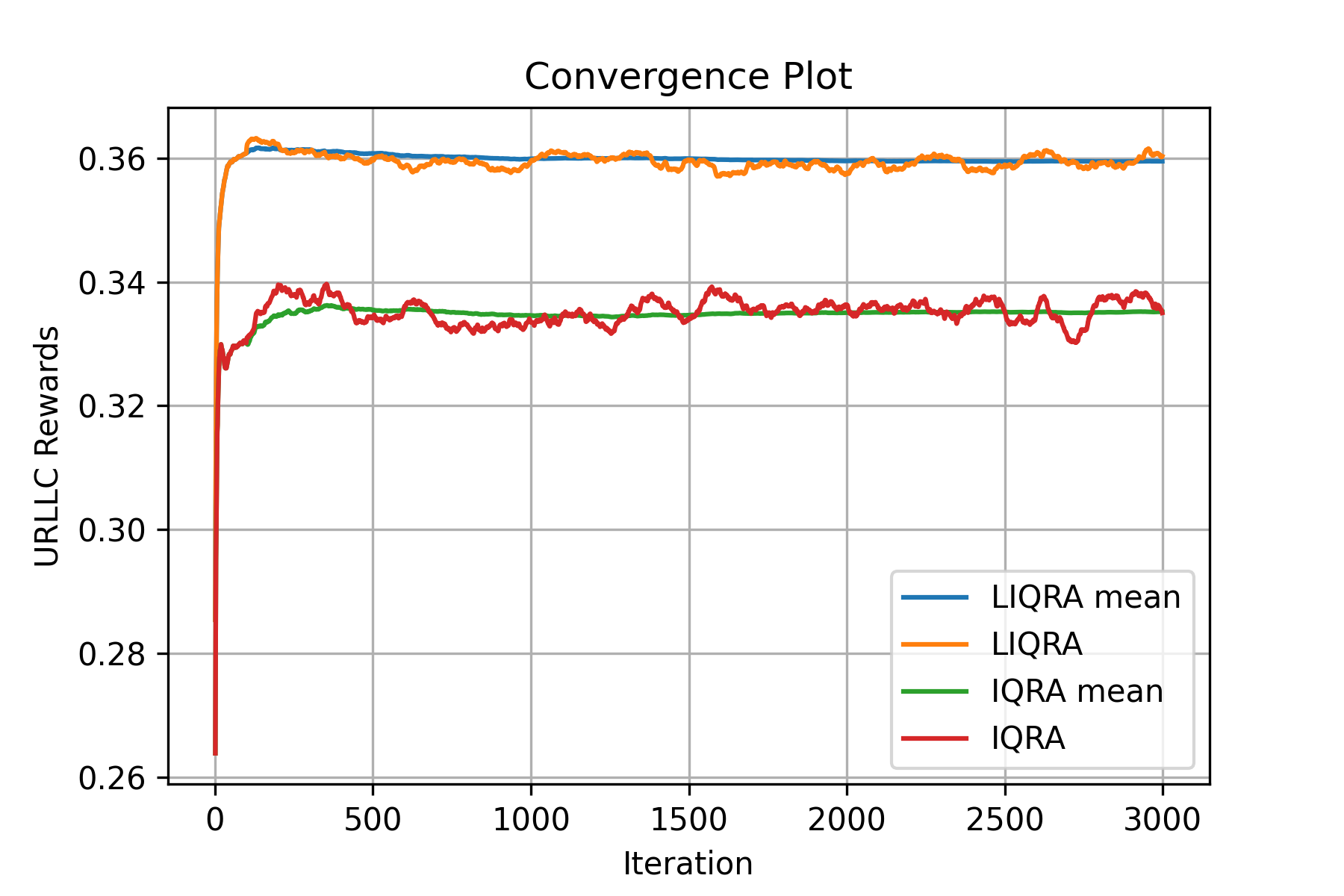}
        \caption{URLLC reward performance for DRL Intelligent agent}
        \label{subfig:subfigure2}
    \end{subfigure}
    \caption{Reward Performance}
    \label{fig:R}
\end{figure}
The results presented in this section are based on the simulation parameters for the environment and DQN agents that were previously described. The eMBB reward function, $R_E$ uses $\alpha_E = 0.7$. Whereas the URLLC reward function, $R_U$ uses $\alpha_U = 0.4$. Different values of $\alpha_s$ have been tested for both slices. The DRL agents use the reward function defined in equation (23). The algorithms are simulated for 3000 iterations of DRL and 30 seconds of simulation time. The figure \ref{fig:R} shows the reward values of eMBB and URLLC slices for both the IQRA and LIQRA algorithms. Subfigure~\ref{subfig:subfigure1} and \ref{subfig:subfigure2} show the value for reward for each run, where LIQRA and IQRA refer to reward values plotted with a sliding window of 100 iterations, and mean refers to the mean reward value. The reward function depends highly on KPI values such as throughput and delay. The instantaneous KPI values are calculated within each TTI, which is 1 ms in duration, and subsequently, individual reward values vary frequently. Therefore, the mean and sliding values of reward have been plotted to analyze the performance of both algorithms. We can see that convergence is achieved within the first 200–300 iterations of DQN. As seen in Subfigure~\ref{subfig:subfigure1} and \ref{subfig:subfigure2}, IQRA performs better for eMBB, and LIQRA has better reward values with URLLC slice. LIQRA has SNR and $\tau$ as UA parameters, whereas IQRA has an estimate of PRB assignment instead of SNR. This is weighted by $w_{s1}$ and $w_{s2}$ for both algorithms. In IQRA, the number of PRBs towards end users is given as one of the parameters to take an UA decision. This associates users with ORUs who can assign more PRBs, achieving higher throughput. The eMBB slice has more stringent throughput thresholds; hence, the more PRBs assigned to end users, the better throughput is achieved, resulting in better reward values. Whereas, for URLLC, latency requirements are more strict, with the first parameters being only SNR helps the algorithm assign users to ORU with better signal quality, giving more importance to the $\tau$ metric.

\begin{figure}[htbp]
    \centering
    
    \begin{subfigure}{0.23\textwidth}
        \centering
        \includegraphics[width=\textwidth]{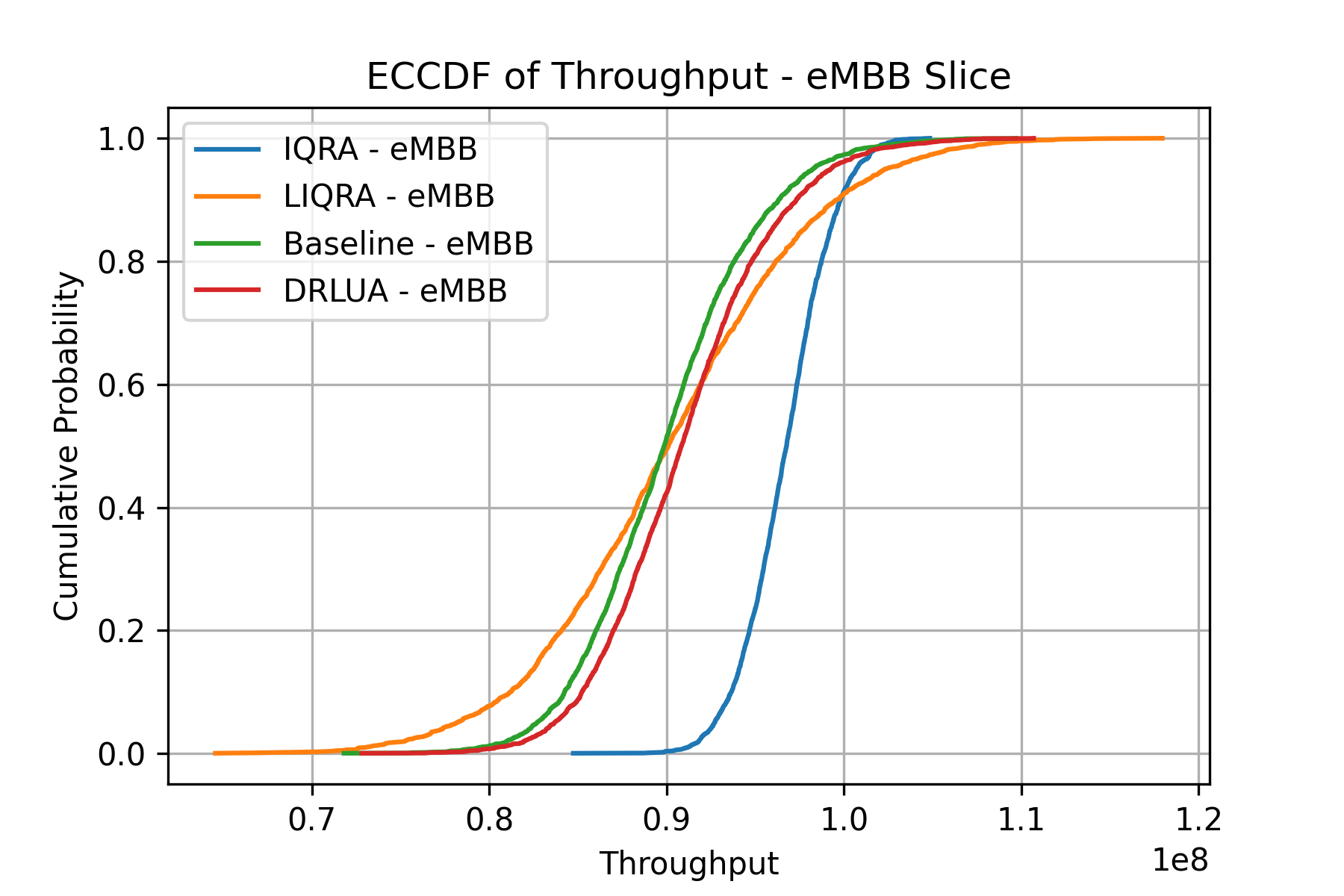}
        \caption{ECCDF of Throughput for eMBB Slice}
        \label{Comp:1}
    \end{subfigure}
    \hfill
    \begin{subfigure}{0.23\textwidth}
        \centering
        \includegraphics[width=\textwidth]{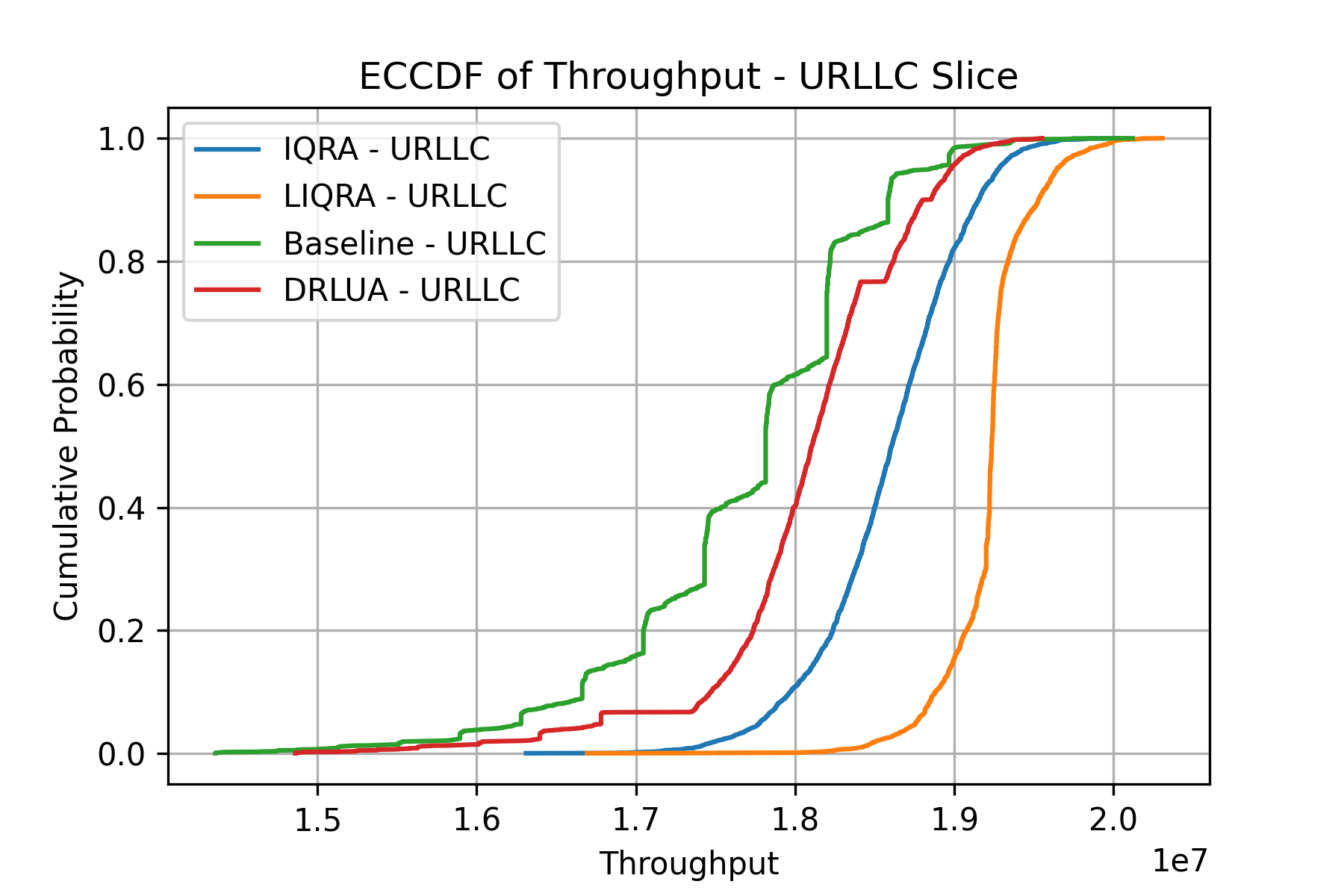}
        \caption{ECCDF of Throughput for URLLC Slice}
        \label{Comp:2}
    \end{subfigure}
    \hfill
    \begin{subfigure}{0.23\textwidth}
        \centering
        \includegraphics[width=\textwidth]{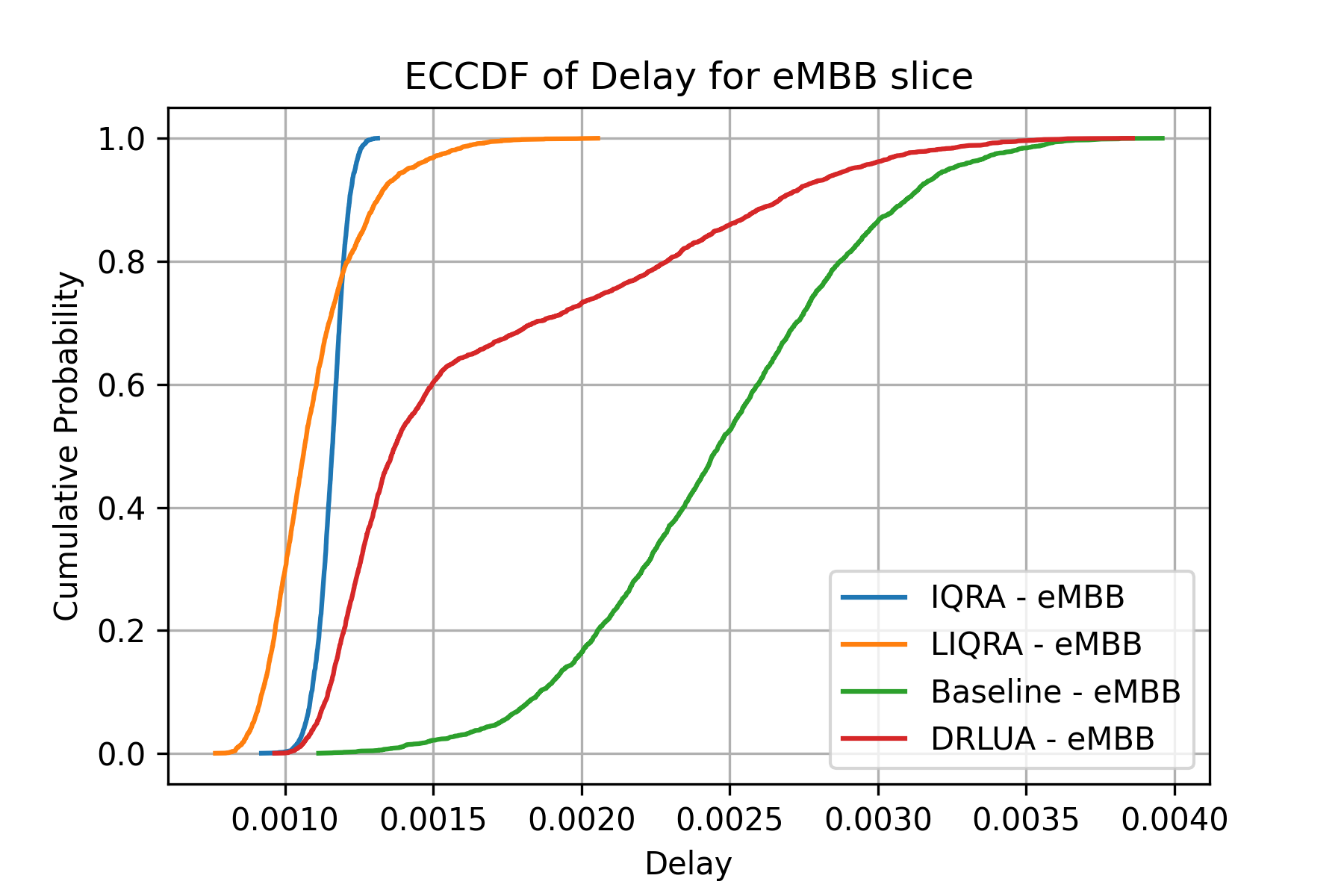}
        \caption{ECCDF of Latency for eMBB Slice}
        \label{Comp:3}
    \end{subfigure}
    \hfill
    \begin{subfigure}{0.23\textwidth}
        \centering
        \includegraphics[width=\textwidth]{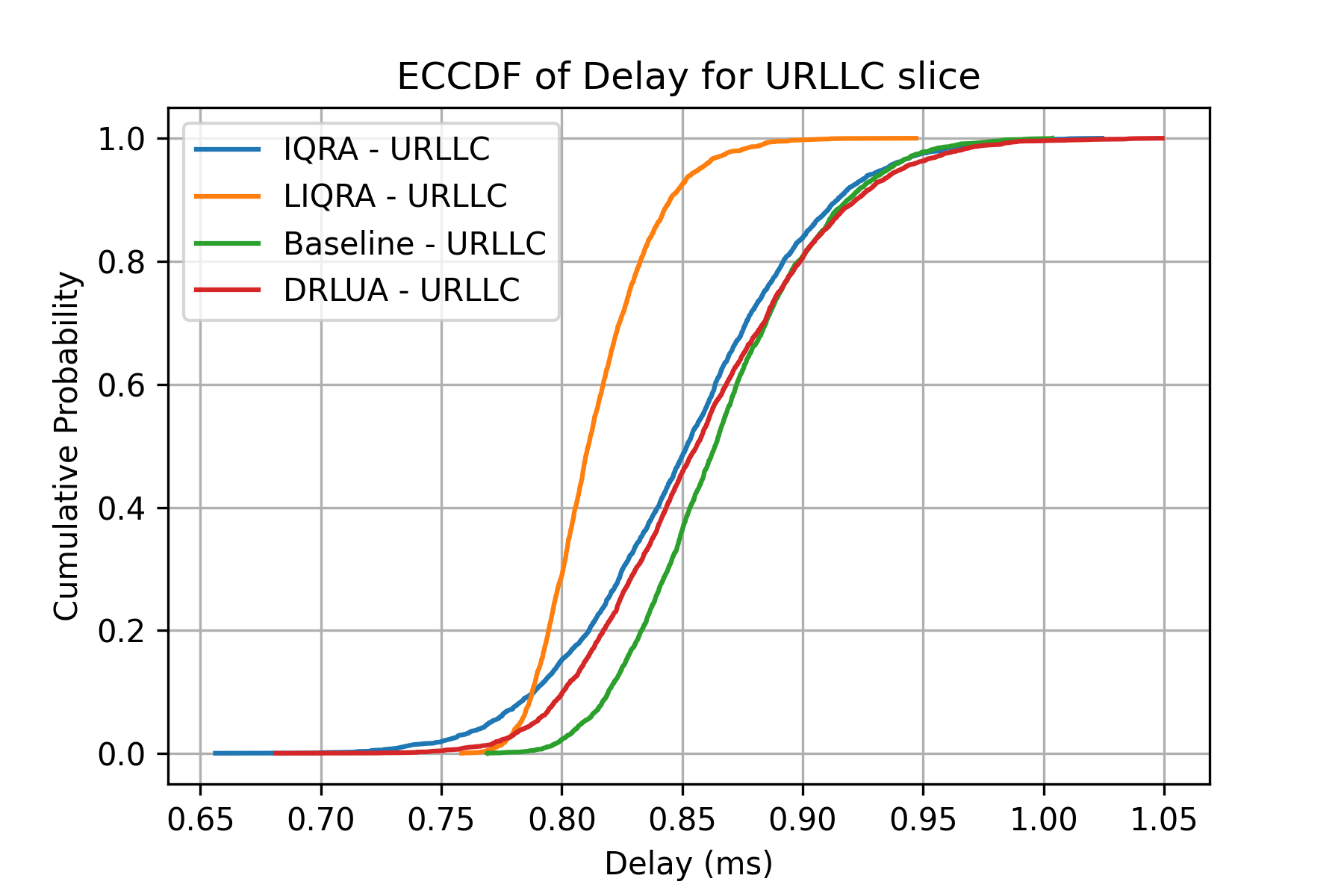}
        \caption{ECCDF of Latency for URLLC Slice}
        \label{Comp:4}
    \end{subfigure}
    \caption{System Performance Comparision}
    \label{fig:Comp}
\end{figure}

Figure \ref{fig:Comp} shows the comparison for system performance between the baseline approach, state-of-the-art (SOTA), referred to as DRLUA now onward, and the proposed algorithms for both slices. The baseline approach uses a simple maxSNR method for user-ORU association, whereas DRLUA uses a number of parallel DNNs approach along with DQN as proposed in \cite{UA}. All of them are implemented on top of proportional fair schedulers. 
Subfigure~\ref{Comp:1} and \ref{Comp:2} show the Empirical Cumulative Distribution Function (ECCDF) for system throughput over DRL iterations for baseline, IQRA, LIQRA, and DRLUA. Both the proposed algorithms outperform the baseline and SOTA approaches. The resource allocation to end users is performed separately for each slice category. Subfigure~\ref{Comp:3} and \ref{Comp:4} show the ECCDF for latency for all four approaches in the simulation. 

The algorithm makes decisions based on weighted values of UA parameters learned from intelligent agents. This allows the proposed methods to customize the importance of decision-making parameters specific to the current state of the wireless environment. For the eMBB slice, IQRA works better among the 4 approaches, as it considers estimates of PRBs and the $\tau$ metric as explained earlier in reward convergence results. In simulation runs, we observe that IQRA selects higher values for $w_{s1}$ compared to $w_{s2}$ in the majority of cases for eMBB slice. Hence, users are associated with ORUs by giving more importance to the quantity of available resources for transmission considering the allowed delay threshold for eMBB is larger, up to 10 ms. Whereas for URLLC, based on the network state, the importance lies with $\tau$ as the delay thresholds are more strict up to 2 ms. The algorithms allocate more resources to the users based on the QoS thresholds, availability of resources, and state of the network, including the current traffic load. It prioritizes and balances the decisions for UA.

\begin{figure*}[htbp]
    \centering
    
    \begin{subfigure}{0.23\textwidth}
        \centering
        \includegraphics[width=\textwidth]{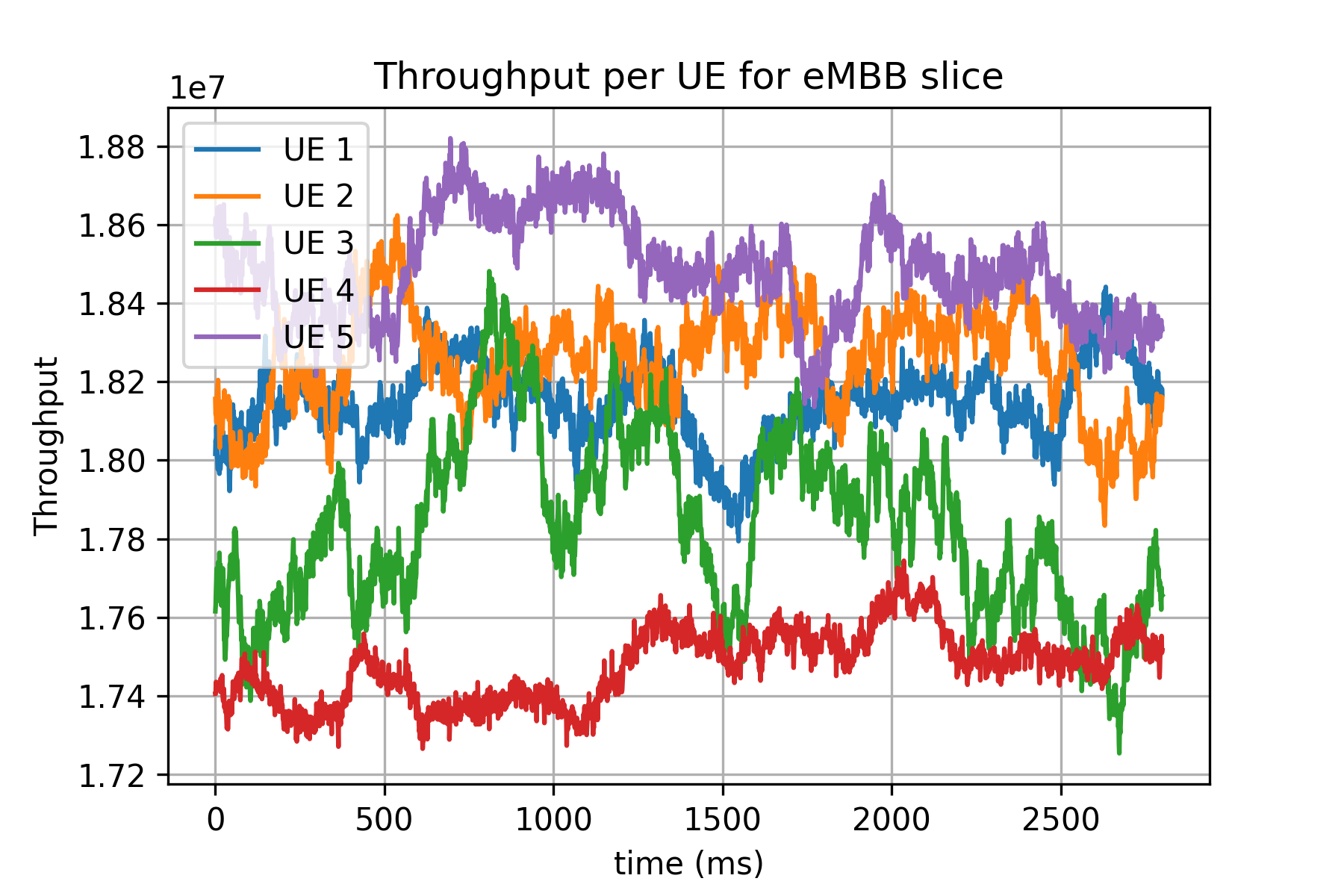}
        \caption{User Throughput eMBB}
        \label{P:1}
    \end{subfigure}
    \hfill
    \begin{subfigure}{0.23\textwidth}
        \centering
        \includegraphics[width=\textwidth]{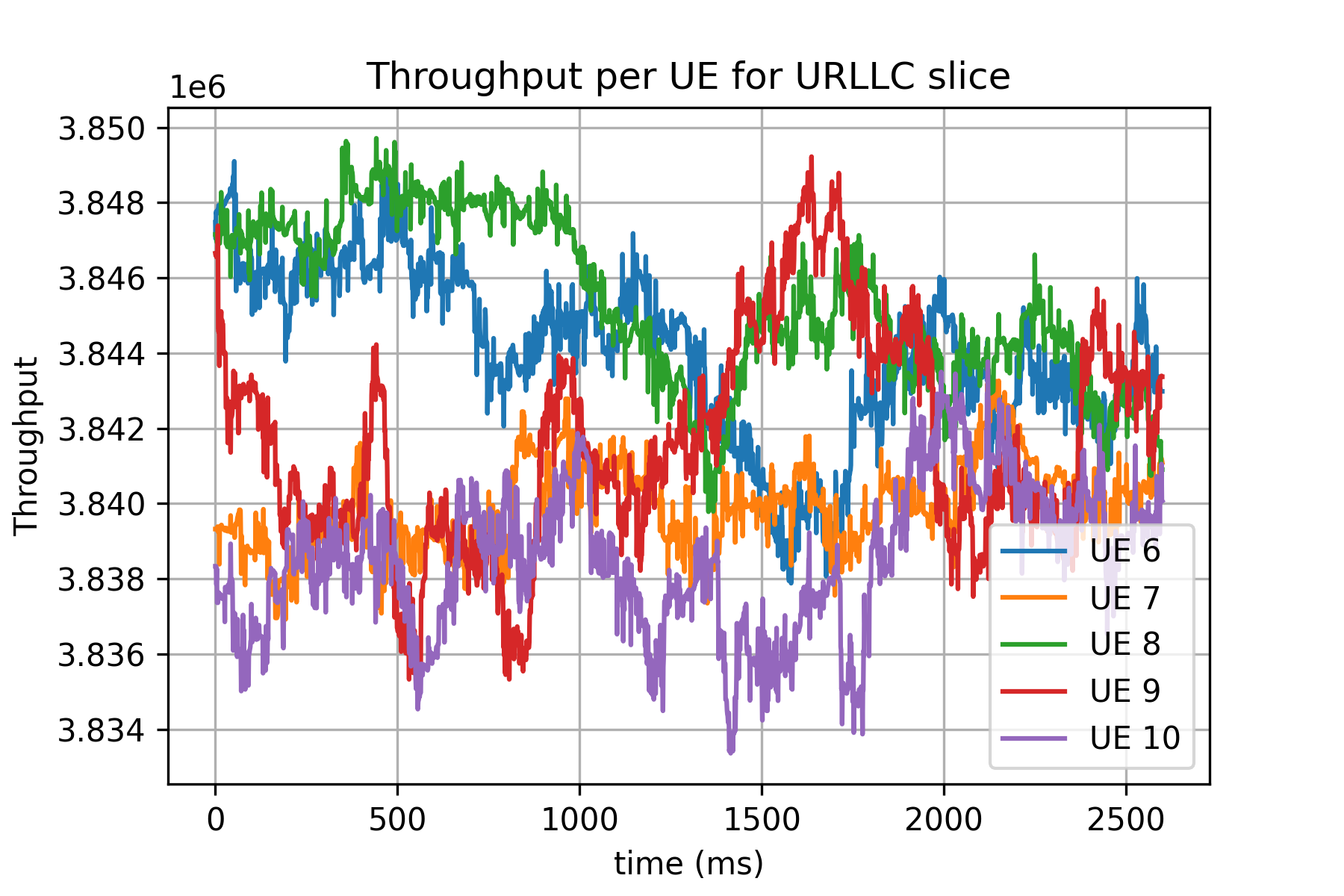}
        \caption{User Throughput URLLC}
        \label{P:2}
    \end{subfigure}
    \hfill
    \begin{subfigure}{0.23\textwidth}
        \centering
        \includegraphics[width=\textwidth]{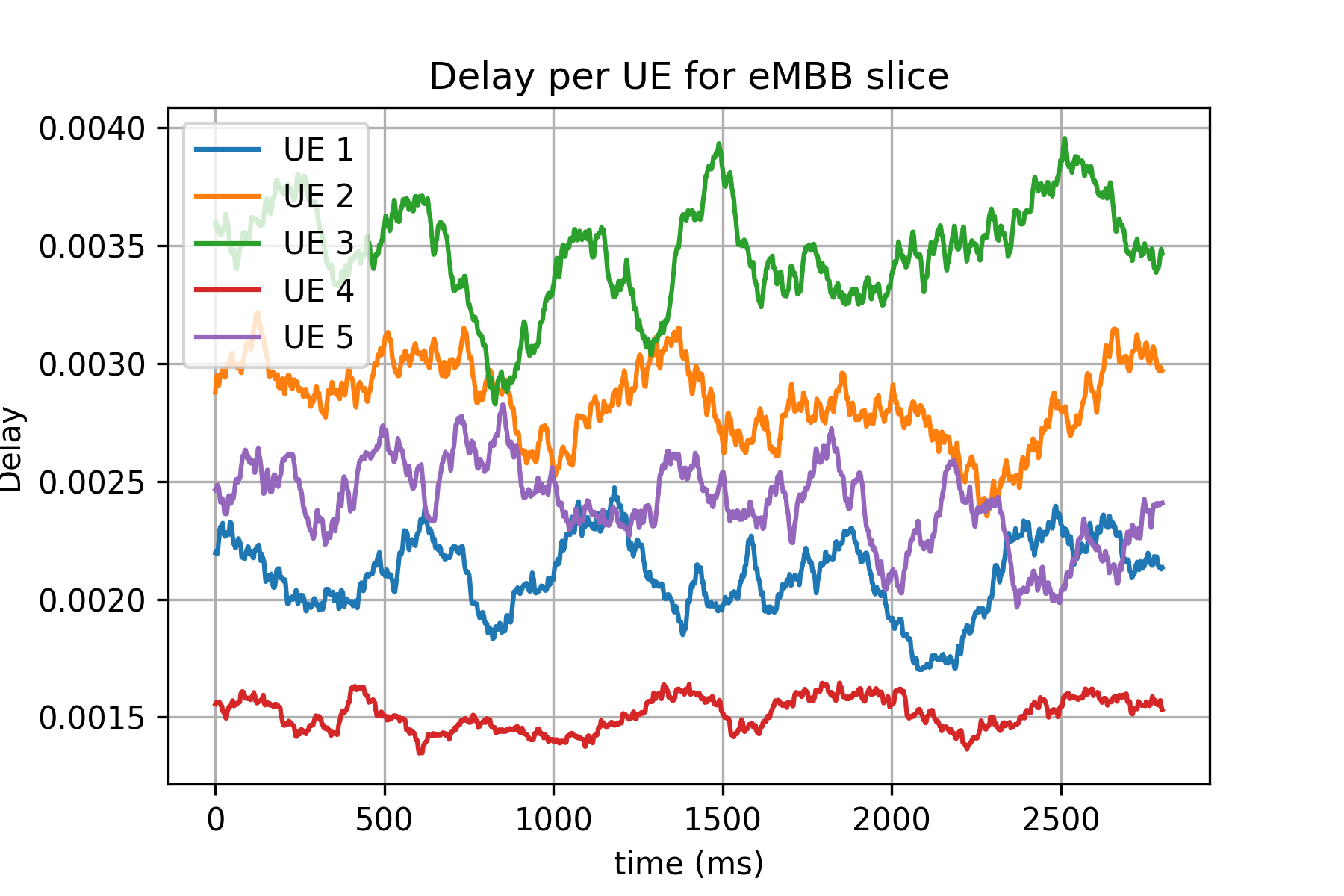}
        \caption{User Delay eMBB}
        \label{P:3}
    \end{subfigure}
    \hfill
    \begin{subfigure}{0.23\textwidth}
        \centering
        \includegraphics[width=\textwidth]{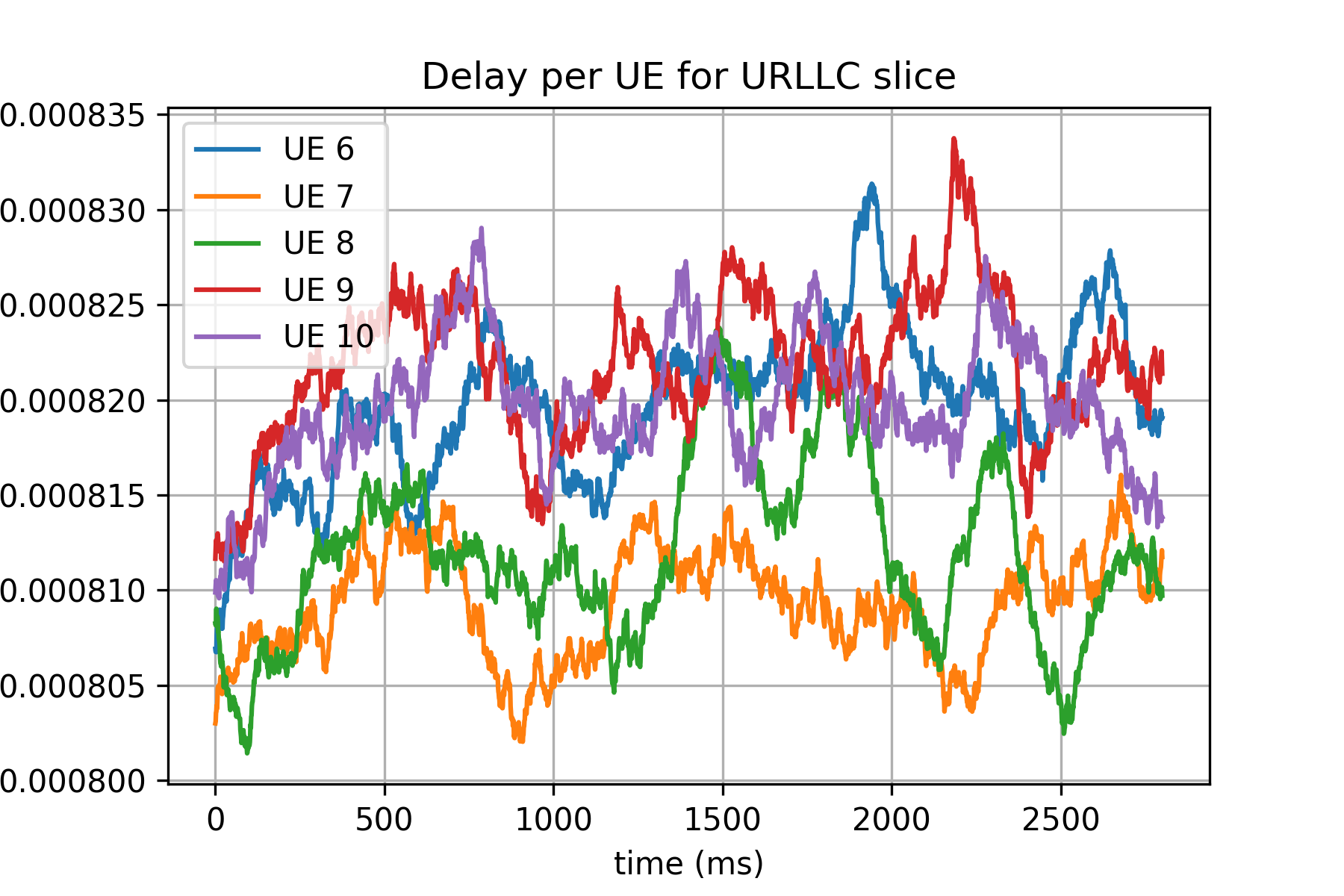}
        \caption{User Delay URLLC}
        \label{P:4}
    \end{subfigure}
    
    \begin{subfigure}{0.23\textwidth}
        \centering
        \includegraphics[width=\textwidth]{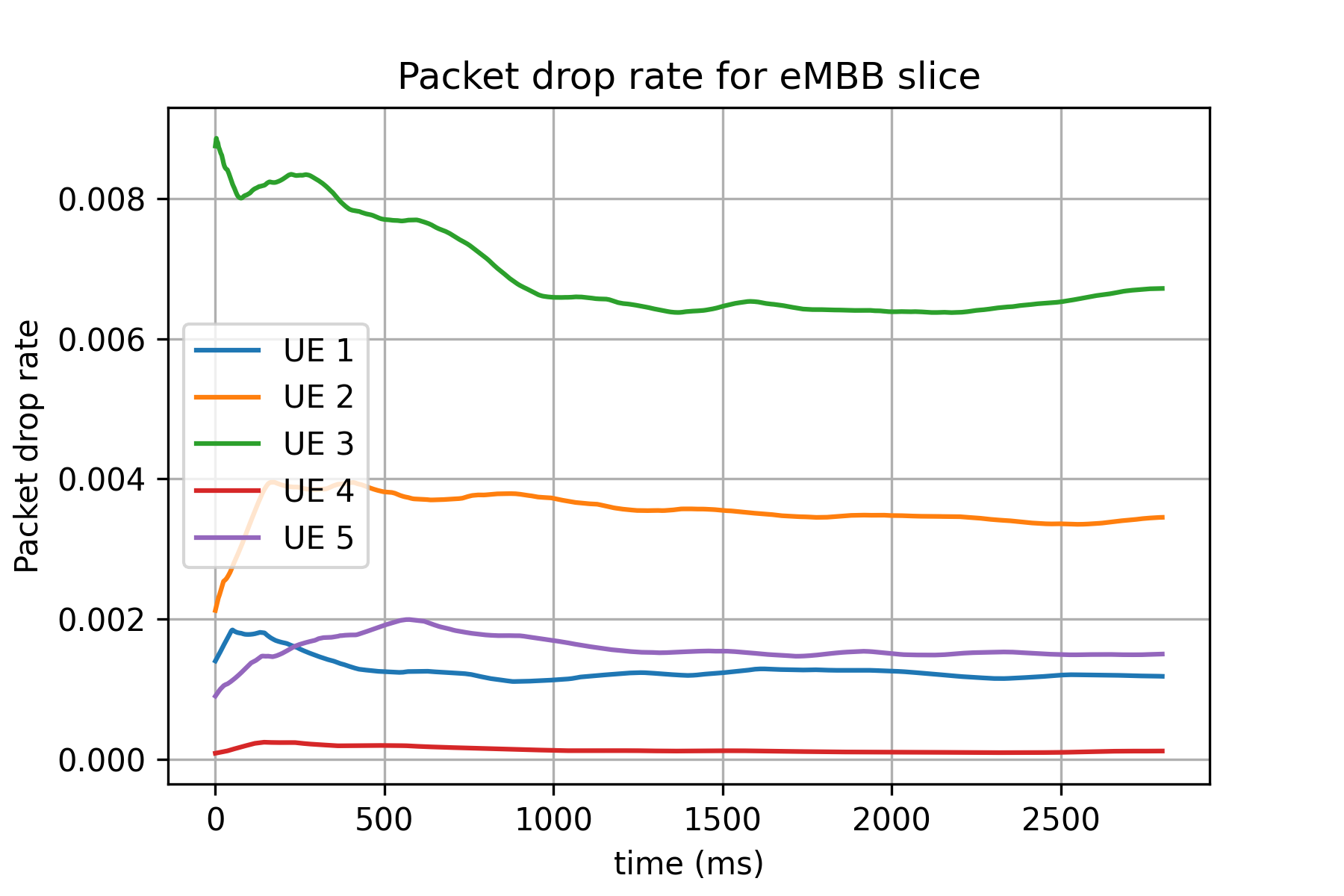}
        \caption{User PDR eMBB}
        \label{P:5}
    \end{subfigure}
    \hfill
    \begin{subfigure}{0.23\textwidth}
        \centering
        \includegraphics[width=\textwidth]{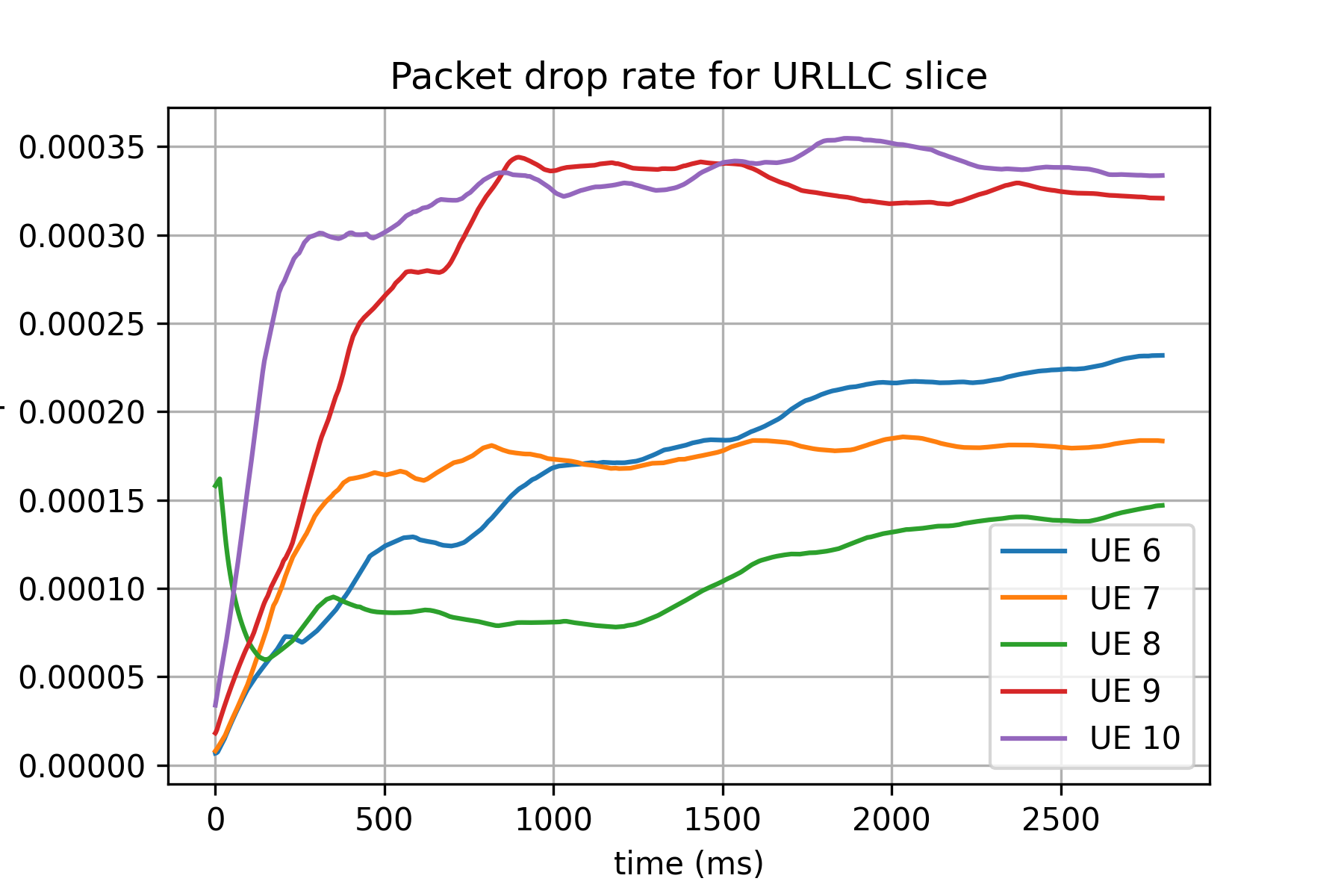}
        \caption{User PDR URLLC}
        \label{P:6}
    \end{subfigure}
    \hfill
    \begin{subfigure}{0.23\textwidth}
        \centering
        \includegraphics[width=\textwidth]{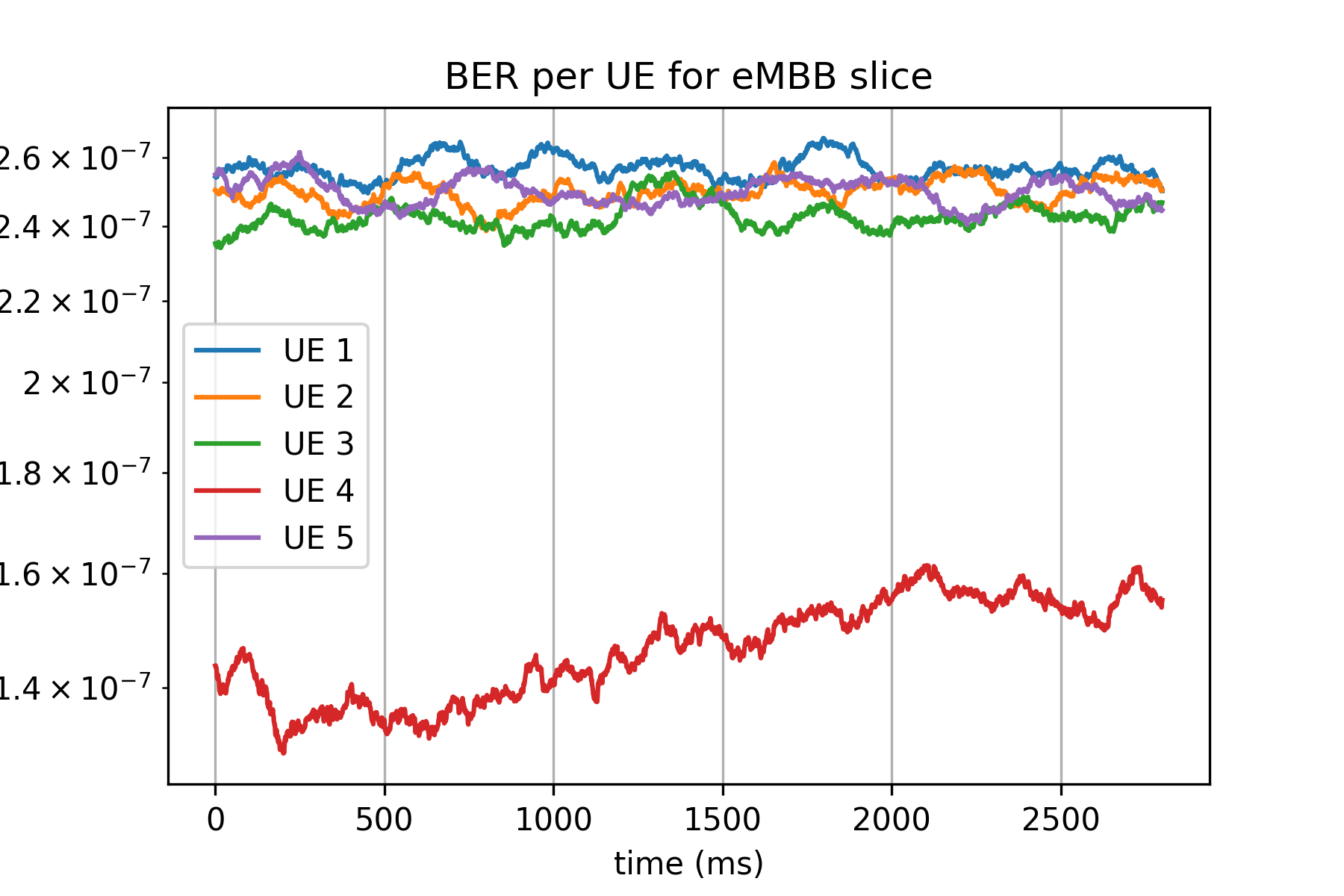}
        \caption{User BER eMBB}
        \label{P:7}
    \end{subfigure}
    \hfill
    \begin{subfigure}{0.23\textwidth}
        \centering
        \includegraphics[width=\textwidth]{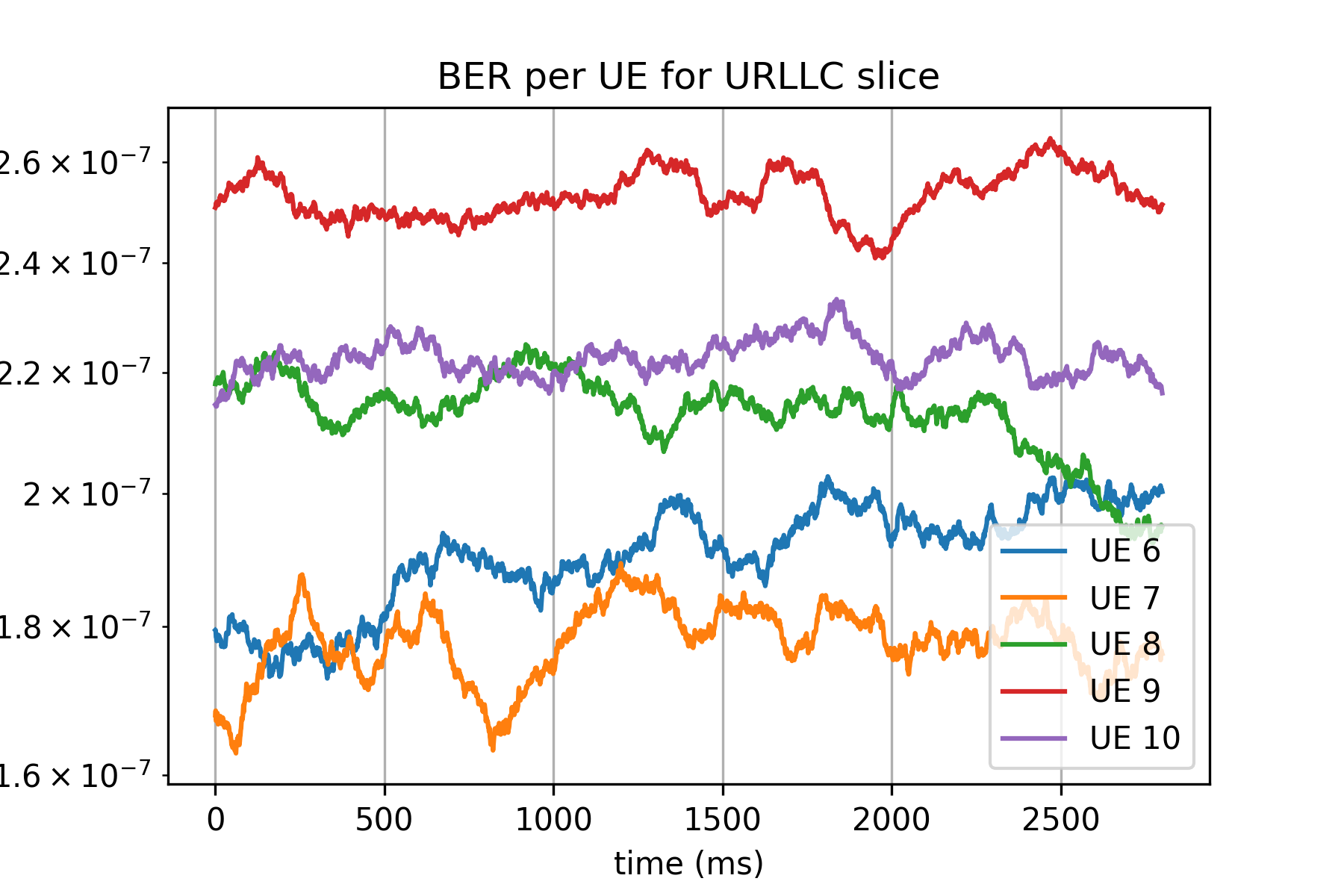}
        \caption{User BER URLLC}
        \label{P:8}
    \end{subfigure}

    \caption{User KPI for eMBB and URLLC slice with IQRA2 Algorithm}
    \label{fig:kpi}
\end{figure*}

Figure \ref{fig:kpi} shows KPI performances, namely, throughput, delay, packet drop rate, and bit error rate for each user under both eMBB and URLLC slices, respectively, using the LIQRA algorithm. Subfigure~\ref{P:1} and \ref{P:2} show experienced throughput for individual users under eMBB and URLLC slices. All users achieve throughput greater than the set QoS thresholds, which are 16 Mbps and 3.8 Mbps for E and U, respectively. Where subfigure~\ref{P:3} shows the delay values for users in eMBB slice. Here, all users experience a delay of less than 4 ms on average for all successfully transmitted packets. Subfigure~\ref{P:5} shows that all users in the eMBB slice have less than 0.5\% packet drop rate. Only 3rd users experience an 0.9\% packet drop rate, which is reduced to 0.7\% as DRL agents make better association decisions. Here, all users experience PDR less than 1\% which is acceptable performance for eMBB services. Further, Subfigure~\ref{P:4} shows all users under URLLC slice experience an average delay less than 1 ms. The QoS threshold for URLLC, $d^{max}_k$ is 2 ms; hence, users are associated with the respective ORU by giving more weight to $\tau_m$ which ensures the available buffer traffic at each UE can be served within the delay budget. Subfigure~\ref{P:6} shows that majority users experience PDR less than $1e^{-4}$. Subfigure~\ref{P:7} and \ref{P:8} give insights about BER performance for users, which is less than $1e^{-6}$ acceptable for all services considered under both slices.

\section{Conclusion}
From the results presented in the above section, we can conclude that the designed intelligent agents and proposed algorithms improved the performance of intra-slice resource allocation compared to baseline and SOTA techniques. The intelligence agent learns different weights $w_{s1}$ and $w_{s2}$ for each slice for the respective parameters affecting association decisions. From the comparison, we can see the proposed IQRA algorithm is more suitable for eMBB, whereas LIQRA performs better for URLLC slices. IQRA provides 11.5\% and 7.42\% improvement in throughput for eMBB slices compared to baseline and SOTA, respectively. While LIQRA provides 19.94\% and 16.54\% improvement in latency compared to baseline and SOTA approaches, respectively. LIQRA improves latency by achieving a minimum latency value of 45.5\% less than compared to the baseline approach. It also improves system throughput for eMBB slices by 6.7\%. Whereas, IQRA improves throughput by achieving an increment in maximum throughput value up to 26.7\% compared to the baseline approach. It also improves the minimum latency value to less than 8.2\% compared to baseline.

\section*{Acknowledgment}
The above work is conducted as a part of MSCA SEMANTIC project  with Innovative Training Networks (ITN) under the grant 861165.   

\ifCLASSOPTIONcaptionsoff
  \newpage
\fi

\begin{IEEEbiographynophoto}{Suvidha Mhatre} received B.Eng. in electronics and telecommunications from Mumbai University, India, M.Sc. in Communication Engineering from Hochschule Darmstadt, Germany and M.Tech. in the same filed from VIT, Vellore, India in 2016, 2019, and 2019 respectively. She is currently pursuing her Ph.D. degree with the Polytechnic University of Catalonia (UPC). She has been an early-stage researcher with Iquadrat Informatica S.L., Barcelona, Spain, working under the MSCA SEMANTIC project in the ITN framework funded by E.C. She has worked with and contributed to multiple European H2020 projects, namely, EuWireless, 5GESSENCE, MORPHEMIC. She has one patent (US20220173849A1) and 2 conference papers published. Her work mainly focus on resource allocation and management using AI/ML techniques in network slicing for ORAN-based 6G networks. e-mail: (suvidha.sudhakar.mhatre@upc.edu).
\end{IEEEbiographynophoto}

\begin{IEEEbiographynophoto}{Ferran Adelantado} is an Associate Professor at the Computer Science, Telecommunications and Multimedia Department and Senior Researcher at the Wireless Networks research lab (WINE) at the Universitat Oberta de Catalunya (UOC). He received the degree in Telecommunication Engineering (2001) and the PhD (2007) from the Universitat Politècnica de Catalunya (UPC), and the BSc in Business Sciences (2014) from Universitat Oberta de Catalunya (UOC). He is also adjunct professor at the Universitat de Barcelona (UB). He has been involved in several National and International projects in the field of wireless communications, and has published more than 60 papers in high impact conferences and journals, and one patent. He served as a TPC and session chair in prestigious International conferences (e.g. IEEE Globecom, IEEE ICC, IEEE PIMRC, etc). e-mail: (ferranadelantado@uoc.edu).
\end{IEEEbiographynophoto}

\begin{IEEEbiographynophoto}{Kostas Ramantas} received the Diploma in computer engineering, the M.Sc. degree in computer science, and the Ph.D. degree from the University of Patras, Greece, in 2006, 2008, and 2012, respectively. In June 2013, he joined IQUADRAT as a Senior Researcher and has co-supervised one Ph.D. student, two ESRs in ITNs, and many secondments in the framework of RISE Projects (e.g., CASPER and WATER4CITIES) and ITN Projects (e.g., Spotlight and 5GAura). He was involved in multiple E.C. Funded Projects (e.g., IoSENSE), while he is currently the TM with 5GMediaHUB Project. He has published more than 35 journal and conference papers. His research interests are in modeling and simulation of network protocols; and scheduling algorithms for QoS provisioning. He was a recipient of two national scholarships. email: (kramantas@iquadrat.com).
\end{IEEEbiographynophoto}

\begin{IEEEbiographynophoto}{Christos Verikoukis} received the Ph.D. degree from the Technical University of Catalonia (UPC), Barcelona, Spain, in 2000. He is currently the Scientific Director with Iquadrat Informatica, and an Associate Proferssor with the Univerisity of Patras. He has authored 151 journal articles and over 200 conference papers. He is the coauthor of three books, 14 chapters in other books, and two patents. He has participated in more than 40 competitive projects. He has served as the Principal Investigator for national projects in Greece and Spain. He is currently the IEEE ComSoc GITC Vice-Chair and the Editorin-Chief of the IEEE NETWORKING LETTERS. e-mail: (cveri@ceid.upatras.gr).
\end{IEEEbiographynophoto}

\end{document}